\documentclass[apj,twocolumn,twocolappendix,numberedappendix]{openjournal}
\usepackage{newtxtext,newtxmath,enumitem,multirow}
\usepackage[utf8]{inputenc}
\usepackage[T1]{fontenc}
\usepackage[breaklinks,colorlinks,allcolors=blue,citecolor=blue,urlcolor=blue]{hyperref}
\usepackage{orcidlink}
\usepackage{ctable}
\pdfoutput=1

\newcommand{\emunit}{\mathrm{cm}^{-6}\,\mathrm{pc}}

\shorttitle{X-rays in the Local Bubble}
\shortauthors{Girichidis, Rea et al.}

\begin{document}

\title{Shaping the diffuse X-ray sky: Structure, Variability and Visibility}

\author{Philipp Girichidis\,\orcidlink{0000-0002-9300-9914}$^{1,\star,\star\star}$}
\author{Erika Rea\,\orcidlink{0009-0009-4849-9764}$^{1,\star,\star\star}$}
\author{Ralf S. Klessen\,\orcidlink{}$^{1,2,3,4}$}
\author{Michael~C.~H.~Yeung\,\orcidlink{0000-0002-5697-0001}$^{5}$}
\author{Efrem Maconi$^6$}
\author{Manami Sasaki\,\orcidlink{0000-0001-5302-1866}$^7$}
\author{Michael Freyberg\,\orcidlink{0000-0001-8158-4631}$^5$}
\author{Juan D. Soler$^8$}
\affiliation{$^1$Universit\"{a}t Heidelberg, Zentrum f\"{u}r Astronomie, Institut f\"{u}r Theoretische Astrophysik, Albert-Ueberle-Str.\ 2, 69120 Heidelberg, Germany}
\affiliation{$^2$Universit\"{a}t Heidelberg, Interdisziplin\"{a}res Zentrum f\"{u}r Wissenschaftliches Rechnen, Im Neuenheimer Feld 225, 69120 Heidelberg, Germany}
\affiliation{$^3$Harvard-Smithsonian Center for Astrophysics, 60 Garden Street, Cambridge, MA 02138, U.S.A. \label{CfA}}
\affiliation{$^4$Elizabeth S. and Richard M. Cashin Fellow at the Radcliffe Institute for Advanced Studies at Harvard University, 10 Garden Street, Cambridge, MA 02138, U.S.A. \label{Radcliffe}}
\affiliation{$^5$Max-Planck-Institut f\"ur extraterrestrische Physik, Giessenbachstrasse 1, 85748 Garching, Germany}
\affiliation{$^6$University of Vienna, Department of Astrophysics, T\"urkenschanzstraße 17, 1180 Wien, Austria}
\affiliation{$^7$Dr. Karl Remeis Observatory, Erlangen Centre for Astroparticle Physics (ECAP), Friedrich-Alexander-Universität Erlangen-Nürnberg,
Sternwartstraße 7, 96049 Bamberg, Germany}
\affiliation{$^8$Istituto di Astrofisica e Planetologia Spaziali (IAPS). INAF. Via Fosso del Cavaliere 100, 00133 Roma, Italy}
\thanks{$^\star$~~P. Girichidis and E. Rea contributed a comparable amount to this work.}
\thanks{$^{\star\star}$\href{mailto:philipp@girichidis.com}{philipp@girichidis.com}, \href{mailto:rea.erika98@gmail.com}{rea.erika98@gmail.com}}

\begin{abstract}
The Local Bubble (LB) is a hot, low-density cavity in the solar neighborhood, inside which the Solar System is currently located. The X-ray emission from such bubbles is strongly governed by the gas density, temperature, and the effects of line-of-sight column density. Yet the physical processes that control the formation and evolution of this emission remain incompletely understood. We analyze a LB analogue identified within a magnetohydrodynamical simulation to investigate the key physical factors that shape its X-ray properties. In post-processing, we examine the spatial distribution, variability, and observational constraints of the X-ray emission. Our study reveals three main results: (1) Shortly after a supernova (SN), the bulk of the X-ray emission arises from a small fraction of the bubble's volume, concentrated in hot regions around recent SN sites. Approximately 95\% of the X-ray luminosity originates from less than 1\% of the total bubble volume. During quiescent phases without recent SNe, the emission morphology changes substantially, with X-ray-bright regions becoming more volume-filling. (2) Column density effects strongly modulate the observable X-ray signal. Gas with column densities exceeding $N_\mathrm{H} \gtrsim 10^{20} \,\mathrm{cm}^{-2}$ efficiently absorbs soft X-ray photons, limiting the depth to which observations can probe. This absorption causes a significant fraction of the sky to be obscured from external soft X-rays. Differences between active and quiescent phases further influence how much of the total bubble emission is visible from within. (3) The X-ray flux shows pronounced temporal variability on Myr timescales, with SN events producing rapid, transient luminosity enhancements, followed by steep declines due to adiabatic cooling. The total flux varies by several orders of magnitude, with SN-driven peaks fading within $10^5$ years.
\end{abstract}

\keywords{ISM:bubbles, Local Bubble, X-rays, magneto-hydrodynamics} 

\maketitle

\section{Introduction}
\label{sec:intro}

Low-density cavities in the interstellar medium (ISM), often referred to as "superbubbles," are ubiquitous throughout the Milky Way and other galaxies \citep{Heiles1979_HI_shells_and_supershells, PinedaEtAl2023, WatkinsEtAl2023_bubble_evolution_ngc628, SandstromEtAl2023}. These structures are generated primarily by the feedback from massive stars via stellar winds and subsequent supernovae (SNe), which inject energy and momentum into the surrounding ISM \cite[e.g.][]{Rahner2017, Rahner2019}. Superbubbles play a pivotal role in shaping the ISM \citep{Klessen2016}, concentrating and distributing stellar feedback effects \citep{KrumholzEtAl2014, KellerEtAl2014}, and triggering the formation of dense gas and stars in the swept up material of the shells and in between bubbles \citep{Elmegreen2011, Dawson2013}. 

Theoretical models and numerical simulations suggest that magnetic fields significantly influence the expansion and morphology of superbubbles, dictating the direction and extent of their growth \citep{FerriereEtAl1991_superbubbles_magnetic, Tomisaka1998_superbubbles_magnetic_blowout_or_confinement, NtormousiEtAl2017}. At the same time, superbubble expansion can also affect magnetic fields by dragging them along as it grows \citep{Alves_2018}. However, the extent to which magnetic fields regulate the formation, expansion, and interaction of superbubbles with their surroundings remains an open question, motivating further theoretical and observational investigations.

As the Solar System is currently located in the interior of a superbubble \cite[e.g.][]{Zucker2023a}, studying our immediate Galactic environment offers the unique opportunity to characterize bubble properties with very high resolution and physical detail. The Local Bubble (LB) was originally discovered in the 1970s based on soft X-ray observations. The data revealed unusually low absorption in specific X-ray energy bands \citep{SandersEtAl1977_soft_xray} and indicated a hot, rarefied cavity with temperatures of approximately $10^6 \, \mathrm{K}$ \citep{McCammonEtAl1983_soft_X-ray_background}. Using dust extinction measurements for nearby stars with well-determined distances has enabled the reconstruction of the three-dimensional (3D) structure of the local ISM with unprecedented precision out to several kpc from the Sun. \cite[e.g.][]{Gaia_2016,2022Lallement,vergely22, Edenhofer2024}. Adding dust continuum and gas line emission data further contributes to our current understanding of the solar neighborhood \cite[see e.g.][]{Alves2020,Zari2023, Soler2023, Soler2025}. 
The 3D maps show that the LB has approximately an elliptical shape in the Galactic disk midplane with a diameter of around $200 - 300$ pc, extending into a chimney-like structure at higher Galactic latitudes \citep{SfeirEtAl1999_mapping_local_bubble, LallementEtAl2003_3d_mapping_local_bubble, ONeillEtAl2024}. This complex geometry suggests a dynamically evolved structure shaped by multiple stellar wind bubbles and subsequent SN events.

The LB is estimated to have formed approximately $13-15\,\mathrm{Myr}$ ago from the cumulative effects of approximately $10-20$ SNe originating in the Sco-Cen stellar association \citep{Fuchs2006,ZuckerEtAl2022_local_bubble, SchulreichEtAl2023_radioisotopic_signatures_earth_local_bubble_ii} Interestingly, the Sun entered the LB about $5 \, \mathrm{Myr}$ ago, as suggested by kinematic studies of the surrounding ISM. Radioisotope analyses in Earth's sediment cores indicate nearby SN activity during this period \citep{SchulreichEtAl2017_radioisotopic_signatures_earth_local_bubble_i, SchulreichEtAl2023_radioisotopic_signatures_earth_local_bubble_ii}.
The surrounding shell of swept-up material has a mass of $\sim 6\times 10^5\,$M$_\odot$ \citep{ONeillEtAl2024} and is an active site of stellar birth \cite[see e.g.][]{ZuckerEtAl2022_local_bubble}, linking the dynamics of the LB to nearby molecular clouds and young star clusters \cite[][]{Zucker2021,Swiggum2024,Cahlon2024A}. Polarization observations also indicate strong magnetic fields accumulated in the LB walls  \citep{PelgrimsEtAl2020, ONeil2025}.

The LB's unique characteristics have been extensively studied through X-ray observations, which reveal important details about its physical properties. The plasma temperature, a defining feature of the LB, is typically around $10^6\,\mathrm{K}\approx 0.1\,\mathrm{keV}$ \citep[e.g.][]{snowden97,mccammon02,YeungEtAl2023,Yeung_2024}. However, spatial variations exist, with the southern Galactic hemisphere being hotter ($0.12\,\mathrm{keV}$ or $1.39\times10^6\,\mathrm{K}$) than the northern hemisphere ($0.10\,\mathrm{keV}$ or $1.16\times10^6\,\mathrm{K}$), as observed by SRG/eROSITA All-Sky Survey (eRASS) data \citep{Yeung_2024}.

Electron densities within the LB are remarkably consistent across sightlines, with typical values of $n_e \sim 4 \times 10^{-3} \, \mathrm{cm}^{-3}$ \citep{SnowdenEtAl2014, Yeung_2024}. These densities, combined with emission measure (EM) and column densities, offer critical simulation benchmarks. Observational EM values, such as those reported by \citet{2017ApJ...834...33L} and \citet{Yeung_2024}, align with regions of low neutral hydrogen column density, highlighting the interplay between hot plasma and surrounding dense gas.

Despite significant advances, challenges remain in isolating the LB's intrinsic properties due to contamination primarily from the solar wind charge exchange (SWCX) process \citep[see a review from][]{Kuntz19}. \citet{2017ApJ...834...33L} combine the results from the {\it DXL} mission to quantify and remove the SWCX contribution in the R1 and R2 (R12) bands in the ROSAT All-Sky Survey taken at solar maximum condition, in which the remaining R12 counts are attributed to the LB emission. More recently, improved characterization of the SWCX process has arrived in the form of the eRASS data: the four completed eRASSs are the first X-ray surveys to be unaffected by the highly variable geocoronal SWCX from the Earth's exosphere due to SRG/eROSITA's halo orbit around L2, and the repeating nature of the eRASSs enables the monitoring of the heliospheric SWCX contribution. \citet{YeungEtAl2023} select the sight lines towards giant molecular clouds located on the wall of the LB, which isolates the foreground and heliospheric SWCX emissions from the more distant background emission components thanks to the clouds' high column densities. Combining these sight lines with simultaneous spectral fitting of four eRASSs, they show that the SWCX contamination in the first eRASS (eRASS1) is low and then increases monotonically with time, closely following solar activity. Utilizing the fact that SWCX is almost negligible in eRASS1, \citet{Yeung_2024} measure the temperature and EM of the hot gas in the LB over the western Galactic hemisphere.

While observational studies provide critical insights into the LB, simulations are essential for understanding its evolution and the underlying physical processes. The Simulating the Life-Cycle of molecular Clouds (SILCC) project, which models sections of the Galactic disk including SN feedback, gravity, and gas dynamics \citep{WalchEtAl2015_silcc_i, GirichidisEtAl2016b}, offers a promising framework for exploring LB analogues. Recent simulations, such as those by \citet{GirichidisEtAl2018b}, have successfully reproduced the size and structure of the LB, providing a reference for studying its impact on observations performed within it \citep[][]{Maconi_2023,Maconi_2025}.

However, a significant gap remains in our understanding of the LB's X-ray properties. Previous studies have largely focused on its geometric and dynamical characteristics, leaving the mechanisms and evolution of its X-ray emission underexplored. This work aims to bridge that gap by analyzing the X-ray emission from a LB analogue identified within an MHD simulation, investigating how physical conditions such as density and temperature contribute to its emission over time. By comparing simulation results with observational benchmarks, we aim to refine models of superbubble dynamics and the role of stellar feedback in shaping the ISM.

The paper is structured as follows. After the introduction (Section~\ref{sec:intro}), we introduce the numerical simulations and the post-processing technique (Section~\ref{sec:methods}). We then analyze the temporal evolution of the simulated bubble (Section \ref{sec:evo-bubble}), considering the role of key physical processes and their impact over time. Section \ref{sec:simplifications} discusses the modeling assumptions and their limitations, evaluating how they influence the interpretation of our results. Moving toward the observational perspective, we present all-sky maps of the simulated bubble, incorporating column density effects and EM projections (Section \ref{sec:skymaps}). We then analyze the bubble's size in relation to both simulated and observed constraints (Section \ref{sec:bub-size}), followed by a detailed study of the all-sky results, highlighting differences between active and quiescent states and their temporal evolution (Section \ref{sec:results}). Finally, we discuss our findings in a broader context (Section \ref{sec:discussion}) and conclude with our final remarks (Section \ref{sec:conclusions}).

\section{Methods}
\label{sec:methods}
\begin{figure}
  \centering\includegraphics[width=0.48\textwidth]{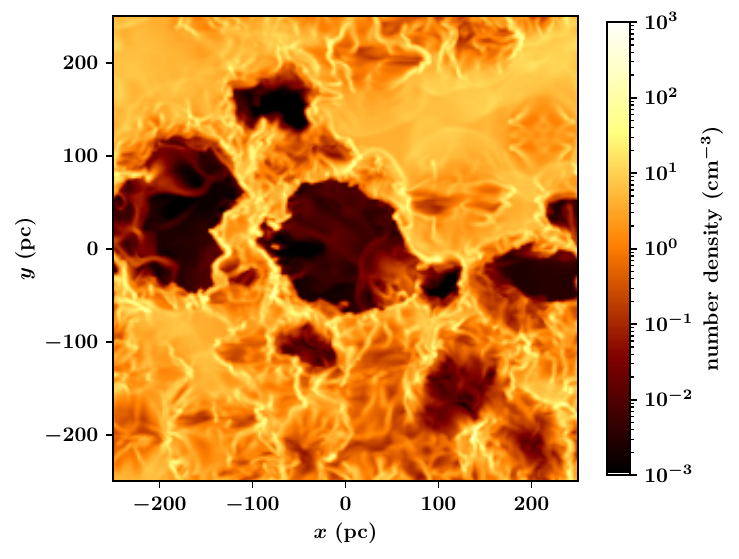}
  \caption{Number density in a cut through the disk midplane of the total simulation box with the selected bubble in the center.}
   \label{fig:illustration-sim-box-and-bubble}
\end{figure}

\begin{figure}
  \centering\includegraphics[width=0.48\textwidth]{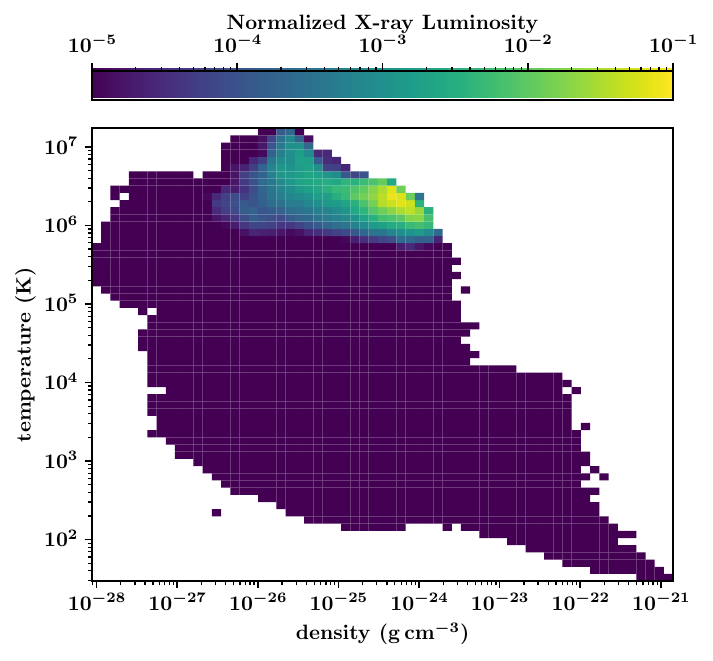}
  \caption{Distribution of emission as a function of temperature and density. Only the hot gas with $T>10^6\,\mathrm{K}$ emits X-rays at relevant intensities. The hot dense gas dominates the emission.}
   \label{fig:2D-histogram}
\end{figure}

\subsection{Simulated LB}\label{sec:sims}

A candidate for the LB is identified within the numerical simulations described by \citet{GirichidisEtAl2018b} and \cite{Girichidis2021}, which are part of the SILCC project \citep{WalchEtAl2015_silcc_i,GirichidisEtAl2016b}. These simulations model a multiphase ISM in a cubic domain of 500\,$\times$\,500\,$\times$\,500\,pc$^{3}$, with physical conditions resembling the local neighborhood. We use the astrophysical code \textsc{Flash} in version 4 \citep{FLASH00,Dubey09} with the split positivity-conserving HLLR 3-wave solver \citet{Bouchut2007, Bouchut2010, Waagan2009, Waagan2011}. The ideal MHD solver ensures divergence-free magnetic fields using \citet{Powell1999}. For the gravitational field, we solve the Poisson equation for self-gravity and add an additional external gravitational field that accounts for the stellar component of the disc. This is modeled as an isothermal sheet \citep{Spitzer1942} with a stellar surface density of $30\,\mathrm{M}_\odot\mathrm{pc}^{-2}$ and a vertical scale height of $100\,\mathrm{pc}$. We use the tree-based method by \citet{WuenschEtAl2018} to compute the total gravitational acceleration.

Stellar feedback is implemented through a constant SN rate. The star formation rate surface density is derived from the Kennicutt-Schmidt relation \citep{KennicuttSchmidt1998} and converted into an SN rate using the initial stellar mass function of \citet{Chabrier2003}. In our simulation box, we inject $10^{51},\mathrm{erg}$ of thermal energy for each of the $15$ SNe occurring per Myr. SN feedback is active from the start of the simulation, and we distinguish between different SN types and spatial distributions \citep{TammannLoefflerSchroeder1994}.

We include a type~Ia component (20\% of all SNe) that is uniformly distributed in the $x$--$y$ plane and follows a Gaussian distribution in $z$ with a scale height of $300,\mathrm{pc}$ \citep{BahcallSoneira1980, Heiles1987}. The remaining 80\% correspond to type~II SNe with a vertical scale height of $90,\mathrm{pc}$. These are further divided into a runaway fraction (60\%) representing isolated explosions and a clustered component (40\%), where SNe are spatially grouped into clusters containing between 7 and 40 explosions \citep{Heiles1987, KennicuttEtAl1989, McKeeWilliams1997, ClarkeOey2002}.

Heating and radiative cooling are coupled to a chemical network that evolves the non-equilibrium abundances of ionized hydrogen (H$^+$), atomic hydrogen (H), molecular hydrogen (H$_2$), singly ionized carbon (C$^+$), and carbon monoxide (CO). The hydrogen chemistry follows the models of \citet{GloverMacLow2007a, GloverMacLow2007b}, as implemented in \citet{MicicEtAl2012}, and is extended with the CO chemistry of \citet{NelsonLanger1997}.

The network includes the formation and destruction of H$^+$ through collisional ionization, ionization by cosmic rays and X-rays, and radiative recombination. The formation of molecular hydrogen follows \citet{Hollenbach89} and accounts for its destruction by cosmic-ray ionization, collisional dissociation in hot gas, and photodissociation in the presence of the interstellar radiation field.

Radiative cooling incorporates emission from the fine-structure lines of C$^+$, O, and Si$^+$, as well as rotational and vibrational transitions of H$_2$ and CO, Lyman-$\alpha$ cooling, and energy exchange between gas and dust \citep{GloverEtAl2010, GloverClark2012b}. At temperatures above $10^4,\mathrm{K}$, we assume collisional ionization equilibrium and adopt the cooling rates from \citet{GnatFerland2012}. Hydrogen, however, is not assumed to be in collisional ionization equilibrium; its cooling contribution is computed self-consistently based on the evolving abundances.

The initial gas distribution follows a stratified Gaussian profile with a scale height of $60\,\mathrm{pc}$, representing a section of the Galactic plane near the solar radius. Initially, the gas is at rest, warm with a temperature of $10^4\,\mathrm{K}$ and in pressure equilibrium. Specifically, they assume a total gas surface density of $\langle\Sigma_\mathrm{gas}\rangle=10\,\mathrm{M}_\odot\,\mathrm{pc}^{-2}$. The magnetic field is initialized along the $x$-direction with a central strength of $B(z=0) = 3\,\mu$G, scaling vertically as $B(z) = B(z=0)(\rho(z)/\rho(z=0))^{1/2}$. During the evolution of the simulation, the central gas number densities, thermal pressures and magnetic field strength evolve to $n=0.1-10\,\mathrm{cm}^{-3}$, $\langle P/k_\mathrm{B}\rangle=1-2\times10^3\,\mathrm{K\,cm}^{-3}$, and $B \sim 0.1-10\,\mu\mathrm{G}$, respectively. We find appropriate chemical compositions using SN rates comparable to the local conditions \citep[see][]{GirichidisEtAl2018b}. The simulations employ adaptive mesh refinement, achieving a maximum resolution of $0.98\,\mathrm{pc}$.

The SN feedback drives turbulence and creates a dynamic three-phase medium akin to the observed local ISM. Individual and clustered SN explosions produce dense filaments and cold gas clouds, primarily in regions where shock shells intersect, while generating hot, low-density voids near clustered explosion sites. We emphasize that the initial conditions are smooth and all density structures, such as clouds, filaments and bubbles, are created by SNe rather than a combination of stellar wind feedback and subsequent SNe.

From the simulation snapshots, we manually identify a series of candidate bubbles at different time steps. For each bubble, we calculate properties such as the average density, total mass of ionized gas, and mass- and volume-weighted magnetic field strength and temperature. This analysis enables us to select a cavity whose properties most closely resemble those of the observed LB. Our final candidate originates from the explosions of 17 clustered SN (see Appendix~\ref{app:SN-positions}), consistent with estimates of $14 - 20$ SN events responsible for the LB \citep{Fuchs2006,ZuckerEtAl2022_local_bubble}. We illustrate the total simulation box and the selected bubble in Fig.~\ref{fig:illustration-sim-box-and-bubble}.

In our simulation, the gas mass swept up by these SN explosions is $M_{\rm{shell}} \, \sim \, 2 \times 10^5\,\mathrm{M}_{\odot}$, which is three times lower than the estimate by \cite{ONeillEtAl2024} and about an order of magnitude smaller than the value suggested by \citet{ZuckerEtAl2022_local_bubble}. However, we emphasize that the total energy driving the cavity's expansion is more critical than the shell mass. The measured shell mass depends on factors such as the integration path and the structure of the surrounding medium. For instance, in a highly structured environment, much of the shell mass resides in small, dense clouds. Notably, the mass estimate in \citet{ZuckerEtAl2022_local_bubble} includes nearby clouds like Taurus and Ophiuchus (private communication), which are absent in our simulation. As a result, the locally concentrated mass occupies only a small solid angle, minimally affecting the broader analysis of the bubble's sky coverage. While our aim is to identify a cavity that best matches the observed LB, we recognize that an exact match is unattainable, see also discussion in \citet{Maconi_2023}.

\begin{figure*}
    \centering
    \includegraphics[width=0.75\textwidth]{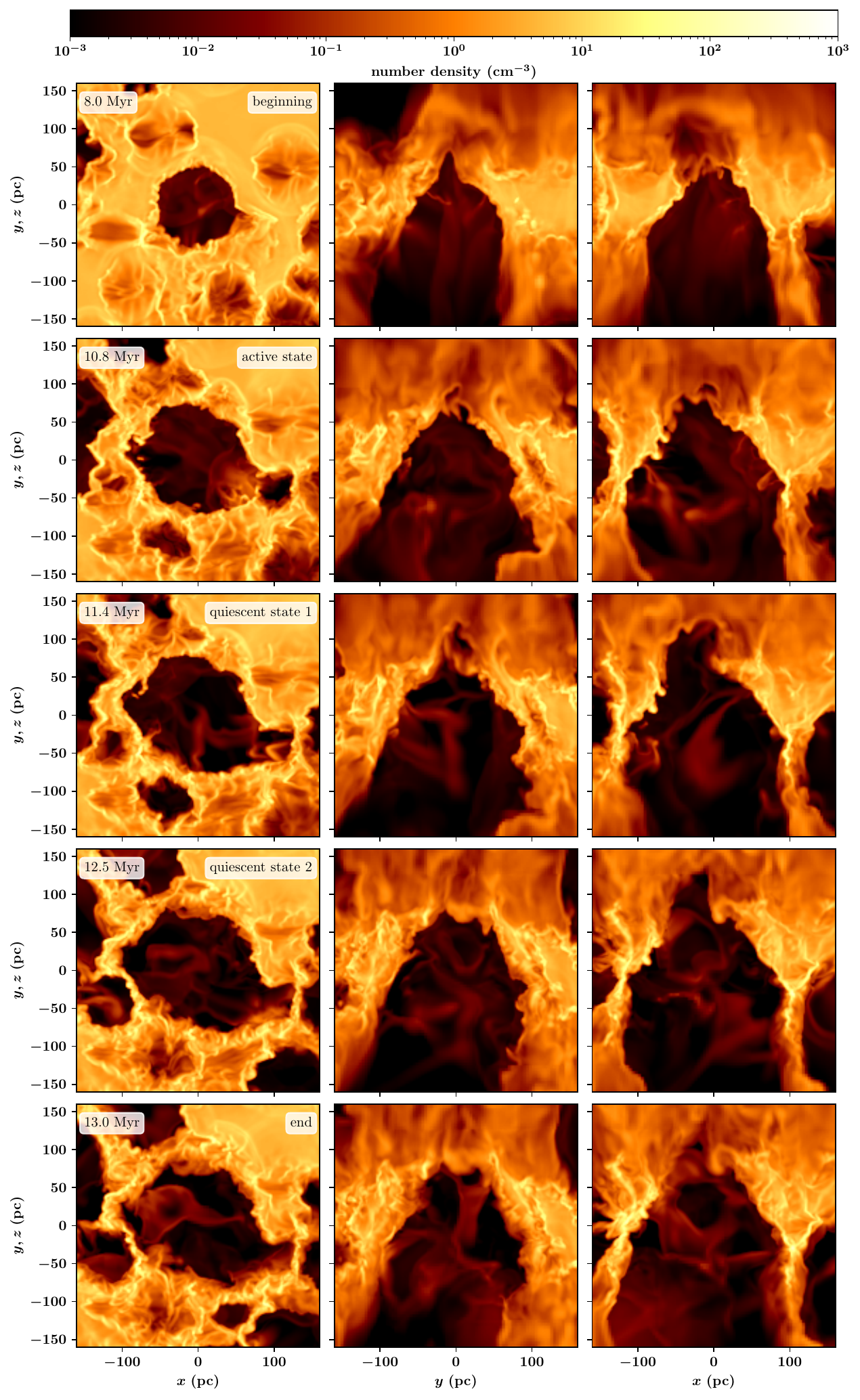}
    \caption{Time evolution of the bubble shown in cuts through the center of the bubble. From top to bottom, we show the different times from $8$ to $13\,\mathrm{Myr}$. The columns correspond to the three different orientations, $x$--$y$ (in the midplane), $y$--$z$, and $x$--$z$ (vertical structure). We note that the bubble is more confined by dense gas at early times (bubble is "closed") and opens up asymmetrically first towards the bottom and later also at the top.}
    \label{fig:bubble-time-evol}
\end{figure*}

\begin{figure*}
    \centering
    \includegraphics[width=0.75\textwidth]{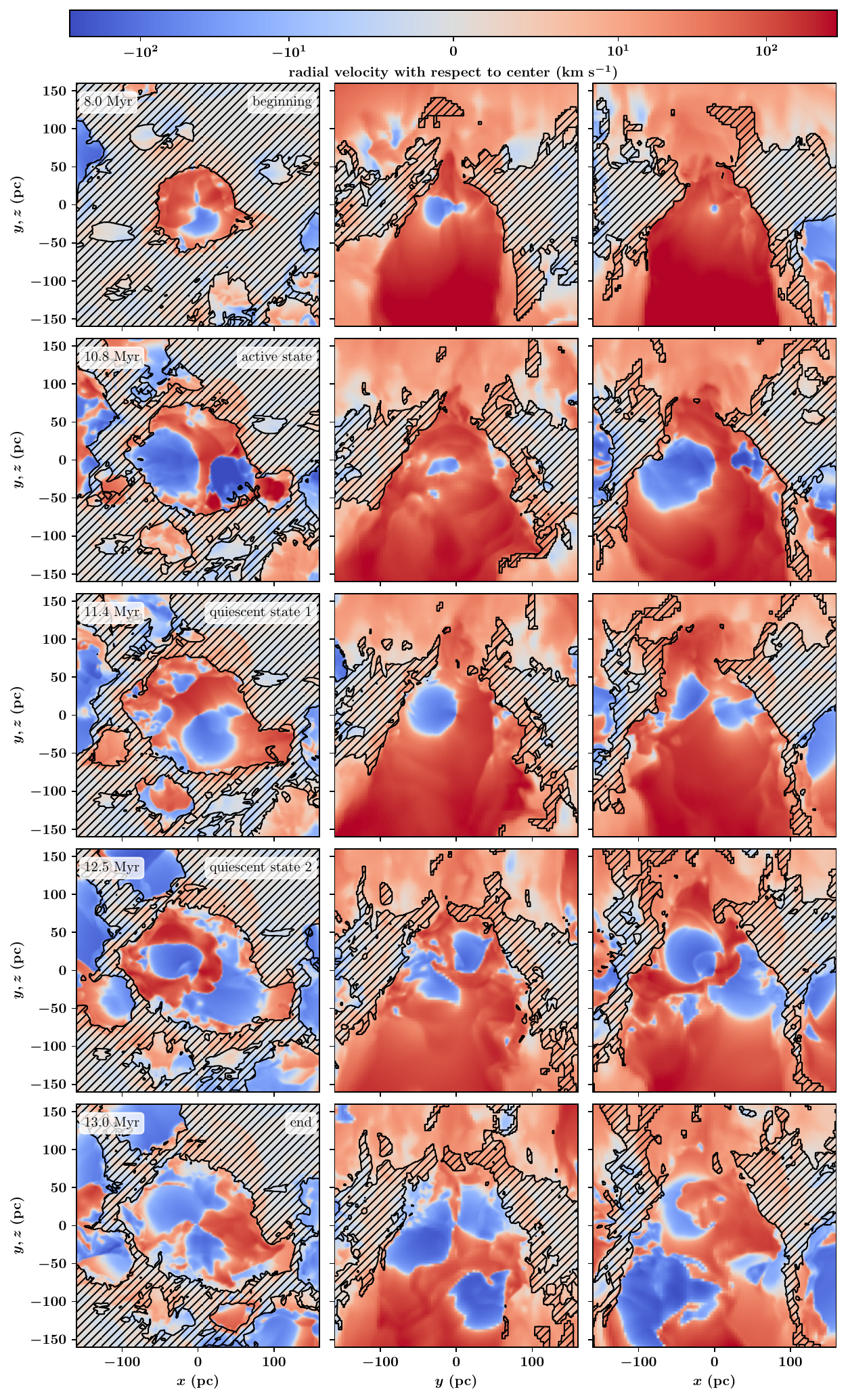}
    \caption{Same as Fig.~\ref{fig:bubble-time-evol} but for the gas dynamics in the simulations. Color-coded is the radial velocity with respect to the center of the bubble, which is the center of the map. The hatched area between the contour lines shows densities $n\gtrsim1\,\mathrm{cm}^{-3}$. We note that the interior of the bubble shows patches of low-density gas moving in different directions with short temporal correlation. The outflow region in the south indicates fast outflowing gas.}
    \label{fig:bubble-time-evol-vrad}
\end{figure*}

\begin{figure*}
    \centering
    \includegraphics[width=0.95\textwidth]{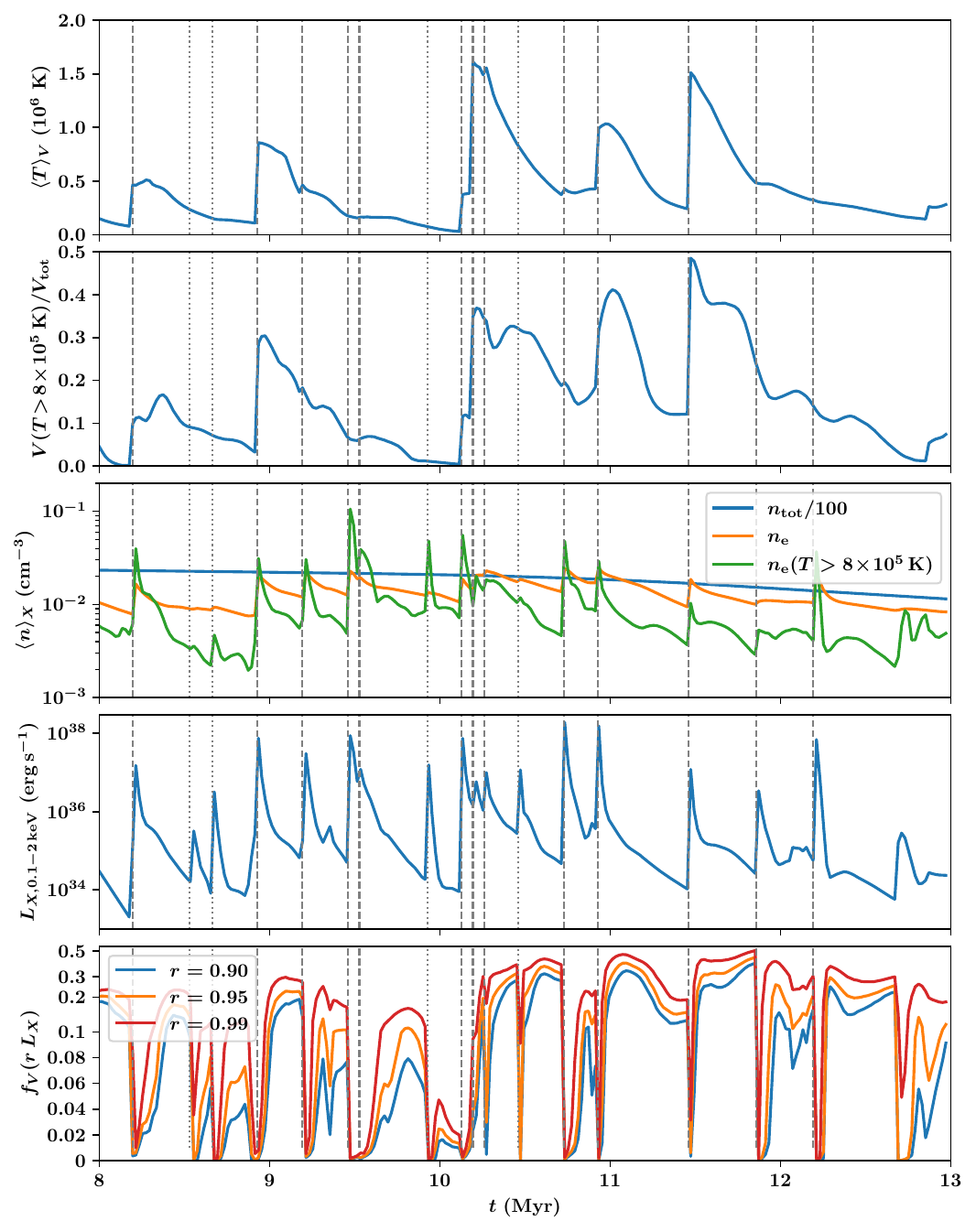}
    \caption{Time evolution of the physical state of the bubble within a 100\,pc radius sphere centered on its origin. From top to bottom the panels show: (1) the volume-weighted temperature, (2) the volume filling fraction of hot gas with $T > 5 \times 10^8$ K, (3) the mean hydrogen number density and the number density of free electrons in the ionized interior (total and from gas with $T > 8 \times 10^5$,K), (4) the total X-ray luminosity, and (5) the cumulative volume fraction responsible for a given fraction of the X-ray luminosity. Vertical dashed lines mark SNe occurring within 100\,pc of the center; dotted lines indicate SNe between 100 and 120\,pc.}
    \label{fig:time_evol-therm-100pc}
\end{figure*}

\begin{figure*}
  \centering
  \includegraphics[width=0.9\textwidth]{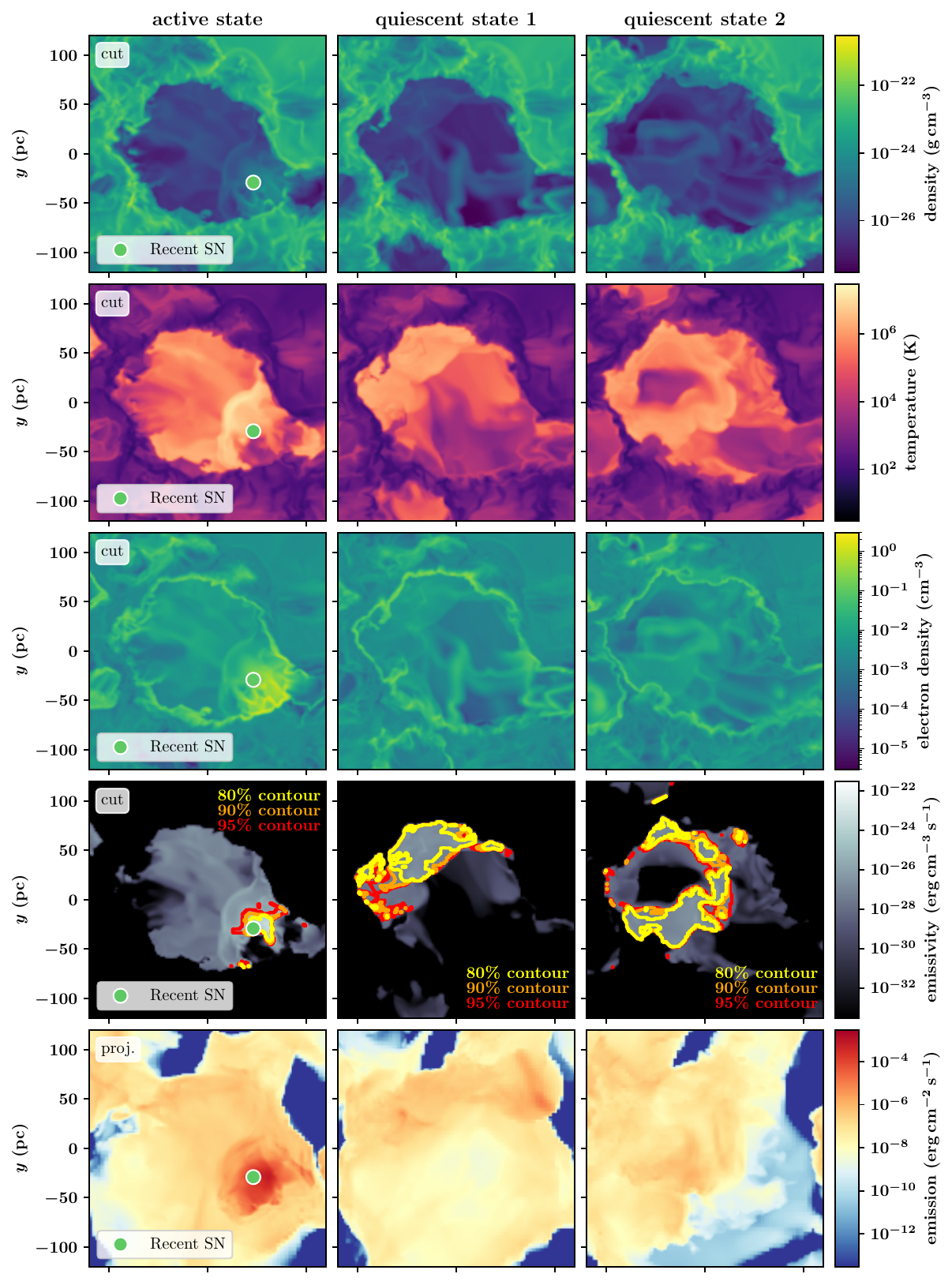}
  \caption{Overview of the simulated bubble at three evolutionary stages. The left column shows an active state shortly after a recent SN explosion ($t = 10.8$ Myr, SN location shown as green dot), while the central and right columns correspond to two quiescent states without recent SN activity ($t = 11.4$ Myr and $t = 12.5$ Myr, respectively). From top to bottom, the first four rows show $x$--$y$ slices through the center of the domain: gas density, temperature, electron density, and X-ray emissivity in the 0.1--2\,keV band (with contours indicating 80\%, 90\%, and 95\% of the total emission). The bottom row displays the corresponding line-of-sight projected X-ray emission.}
   \label{fig:overview-default-bubble}
\end{figure*}

\subsection{X-ray post-processing}
\label{sec:post-processing}

X-rays can be produced via several mechanisms, such as thermal emission, synchrotron radiation, bremsstrahlung and inverse Compton scattering. The X-ray emission spans from $0.1\,\mathrm{keV}$ up to approximately $10\,\mathrm{keV}$ and is categorized into two regimes. \textit{Soft} X-rays, from 0.1 to 2\,keV, are often associated with thermal emissions from hot ($10^6-10^7\,\mathrm{K}$) gas in the ISM or the Galactic halo, as well as emissions from stellar coronae and the early stages of SN remnants. Contrarily, \textit{hard} X-rays are characterized by higher energies, generally between 2 and 10\,keV and beyond. They are typically emitted by more energetic processes, such as those occurring near black holes, neutron stars, and AGNs, where non-thermal processes like synchrotron and inverse Compton scattering dominate. In this paper, we focus on soft X-rays and the thermal emission from a hot ionized plasma.

In order to investigate X-rays in the range of $0.1-2\,\mathrm{keV}$, we post-process the simulations using tabulated emissivities provided by the \textsc{YT} framework (\citealt{TurkEtAl2011}, \href{https://yt-project.org/data/}{YT data page}) using \textsc{Cloudy}\footnote{For \textsc{Cloudy}, see \href{https://gitlab.nublado.org/cloudy/cloudy}{https://gitlab.nublado.org/cloudy/cloudy}.} (\citealt{cloudy2023}) and \textsc{APEC} \footnote{For APEC, see \href{http://www.atomdb.org/}{http://www.atomdb.org/}.}. The tables generated using \textsc{Cloudy} assume a photoionized plasma, whereas the \textsc{APEC} tables use collisionally ionized plasma. Figure~\ref{fig:2D-histogram} illustrates the gas phase with the primary origin of X-rays, where we show the normalized emission in as a function of density and temperature for our bubble at the fiducial time of $t=10.8\,\mathrm{Myr}$. We note that the hot and dense gas emits by far most of the radiation. At temperatures below $T=10^6\,\mathrm{K}$ the emission is negligible. For the density and temperature regime we use, the emissivities of the \textsc{Apec} table are approximately twice as high as for the \textsc{Cloudy} table; we compare the differences in Appendix~\ref{sec:cloudy-vs-apec}. Since the general trends in emission are the same in both tables and our conclusions are the same for both tables, we present results created with the \textsc{APEC} table.

\section{Temporal evolution of the simulated bubble}
\label{sec:evo-bubble}

We show the evolution of the bubble in Fig.~\ref{fig:bubble-time-evol} over the course of the investigation from 8 to 13 $\mathrm{Myr}$. The data shows cuts through the center of the bubble for all three orientations. At early stages, the bubble is surrounded by dense gas in the $x$--$y$ plane. Over time, neighboring bubbles, caused by other clustered SNe, merge into our central bubble and turn the ISM into a more porous medium with molecular clouds located in between the bubbles. In the middle and the right-hand panels we can see the vertical structure. Here we note the asymmetry above and below the midplane. Whereas the dense material above the mid plane can resist the dynamical impact of the SNe, the region below the midplane cannot, which causes the bubble to open up towards the circum-galactic medium. We note that the LB around the Sun appears to be an open Galactic chimney \citep{ONeillEtAl2024}. 

The dynamics inside the bubble is illustrated in Fig.~\ref{fig:bubble-time-evol-vrad}. We show the radial velocity with respect to the center of the bubble. The shaded area indicates densities above $n=1\,\mathrm{cm}^{-3}$. From the velocity maps we can infer several important features in terms of the dynamics. The bubble is dominated by fast flows ($\sim100\,\mathrm{km\,s^{-1}}$) in quickly changing directions. The chimney at the southern part is dominated by fast outflowing gas in all but the last time snapshot. The regions of denser gas show small radially outward pointing velocities (few $\mathrm{km\,s^{-1}}$) reflecting the slow overall expansion of the bubble. Also, we see the velocity structure of neighboring bubbles with negative radial components with respect to the center of the main bubble. At later times, regions of low column density will break up and allow for localizes inflow into the bubble. We illustrate this in Appendix~\ref{sec:mollweide-vrad}.

To get an impression of the thermal evolution, we select a sphere around the center of the bubble with a radius of 100\,pc and measure various key quantities over time. We note that the bubble is not a sphere but has a more complex shape including the open part towards the south. We nonetheless chose a simple spherical investigation of differently weighted quantities in this section and investigate the details of the bubble in the Section~\ref{sec:skymaps}, where we integrate up to a limiting column density from the center of the bubble. The center of the bubble is picked by eye and kept fixed over the temporal evolution. We ensure that quantities that can sensitively depend on the exact location of the center are not influenced by our choice (see Appendix~\ref{sec:fiducial-bubble-center}). In Fig.~\ref{fig:time_evol-therm-100pc} we show the time evolution of the volume weighted temperature, the volume filling fraction of the hot ($T>8\times10^5\,\mathrm{K}$) gas, the total and electron number densities, the total X-ray luminosity in the energy range of order $0.1-2\,\mathrm{keV}$ and the fraction of the volume from which 90, 95 and 99\% of the luminosity originates from. We emphasize that the volume weighted temperature, the electron number density--in particular the one only involving the hot gas--as well as the luminosity predominantly reflect the conditions inside the bubble and are not influenced by the dense shell around the bubble. The grey vertical dashed lines indicate the SNe that occur inside the bubble ($d<100\,\mathrm{pc}$), the grey dotted lines are the SNe that explode in the distance range from $100\le d/\mathrm{pc}\le120$, i.e.\ the SNe at the edge of the shell. There are several important features in the time evolution. First, the sequence of SNe is not equally spaced in time. This is due to the random choice of the SNe and their clustering properties \citep[see][for details]{GirichidisEtAl2016b,GirichidisEtAl2018b}. In many cases, a SN triggers an instantaneous jump in the volume weighted temperature because the SN corresponds to thermal energy injection within a volume that is a noticeable fraction of the bubble volume -- to be precise, it is the volume that encompasses $800\,\mathrm{M}_\odot$ of gas. Depending on the detailed position of the SN within the bubble, the subsequent thermal evolution differs, which manifests in the volume filling fraction of the hot gas: in some cases the thermal injection directly results in an increase in the filling fraction, in other cases there is a temporal delay over which the hot gas expands into the bubble. In addition, external SNe (with a distance larger than 100\,pc from the bubble center) can push gas into the bubble. The details of the position as well as the overlap of SNe with gas of different density also results in a different peak luminosity. We emphasize that the injected energy per SN is $10^{51}\,\mathrm{erg}$ in all cases. Intuitively, one might expect a simple direct scaling of the environmental density, where the denser gas results in more emitting gas and therefore larger luminosities. However, the correlation is more subtle. If SNe explode in lower densities, the gas is likely to be hotter already. The SN energy is therefore more likely sufficient to heat the gas to X-ray emitting temperatures. The same thermal energy injected in dense gas results in overall lower temperatures. Since we inject the SN energy into a sphere, which encompasses a mass of $M=800\,\mathrm{M}_\odot$\footnote{The mass of $M=800\,\mathrm{M}_\odot$ was chosen empirically ensuring that the thermal injection results in hot enough temperatures to properly follow the subsequent expansion.}, the size of this injection region varies depending on the spatial variations of the density. The total luminosity is therefore a complex combination of the average density in the injection region and the local spatial variations. This is further complicated by the fact that we do not store a snapshot at exactly every SN, so the captured peak might not be the true peak, which is visible in some cases, where the vertical line indicating the SNe is offset by a short time. Due to the steep decline shortly after the injection, these small temporal offsets are likely visible.

In the middle panel of Fig.~\ref{fig:time_evol-therm-100pc} we illustrate the time evolution of the number density of free electrons $n_{\rm e}$ in the dilute bubble interior, using orange for the average value and green for the hot gas at temperatures $T>8\times10^5\,\mathrm{K}$. Because the medium in the bubble is fully ionized, the number density of free electrons closely corresponds to the hydrogen number density $n$. Our values of  $n\gtrsim 0.01\,{\mathrm{cm^{-3}}}$ in the bubble interior are in good agreement with the observed values \cite[in the range $0.012 - 0.02,\mathrm{cm}^{-3}$, see e.g.][]{LinskyRedfield2021,YeungEtAl2023,ONeillEtAl2024}. \citet{YeungEtAl2023}, suggests lower values of $0.004\,\mathrm{cm^{-3}}$, assuming the sight line is completely filled with hot plasma. Careful inspection of density field (Figs.~\ \ref{fig:bubble-time-evol} and \ref{fig:overview-default-bubble}) indicates fluctuations of up to a factor of ten around this mean, caused by interacting shocks. This is in line with the presence of low-mass and low-density clouds close to the Sun with $n \approx 0.03 - 0.1\,\mathrm{cm}^{-3}$ \cite[see e.g.][]{Gry2014,Zucker2025}. 

To illustrate the long-term evolution of the local ISM in the simulation, we also plot the mean number density $n_{\rm tot}$ in blue in the middle panel of Fig.~\ref{fig:time_evol-therm-100pc}, which includes the material swept up into the bubble walls. We find values of $1-2\,\mathrm{cm^{-3}}$, about two orders of magnitude higher than in the bubble interior and consistent with the expectations for the average number density in the solar neighborhood \cite[see e.g.][]{Ferriere2001}. We also note that $n_{\rm tot}$  gradually decreases with time as the bubble expands into the surrounding medium. This is because we consider a constant volume over which we take the average. Contrary to the number density in the hot bubble interior, we find no impact of individual SNe on $n_{\rm tot}$, due to the fact that the fast moving hot gas resulting from the energy injection by SNe is very dilute and does not carry noticeable amounts of mass. Therefore, the average number density in the volume only reflects the large-scale and long-term evolution of the region.

\begin{figure}
    \centering\includegraphics[width=0.48\textwidth]{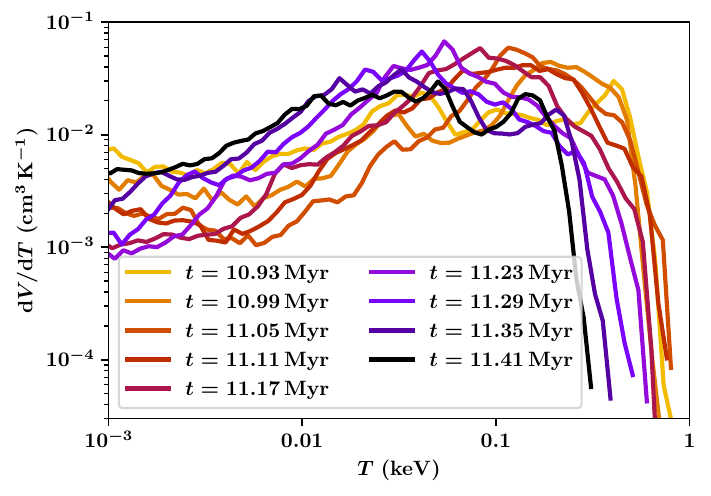}
    \caption{Temperature distribution in the spherical control volume for different times. At the earliest time, a SN just exploded in the bubble. The distributions of $T$ shift by approximately one dex during this phase of $0.5\,\mathrm{Myr}$.}
    \label{fig:pdf-temp-diff-times}
\end{figure}

We note a clear pattern in the time evolution of the luminosity and the emitting volume when looking at the bottom two panels of Fig.~\ref{fig:time_evol-therm-100pc}.  When a SN explodes, the total luminosity increases by several orders of magnitude from $L_X\sim10^{34}$ up to $10^{38}\,\mathrm{erg\,s}^{-1}$. Shortly after the explosion the total luminosity decays. We highlight that this is simply the total value for $L_X$ in the sphere and we investigate further details, including the spatial distribution expressed as X-ray flux maps, in Section \ref{sec:skymaps} below. The fraction of the volume $f_V$ that accounts for most of the X-rays evolves inversely to $L_X$. A peak in the luminosity caused by a recent SN concentrates the emission to the close vicinity of the explosion resulting in  tiny $f_V$ values. The subsequent expansion of the SN and the accompanied cooling of the gas results in much more volume filling emission,  but a lower average X-ray emission. Once the hot gas is distributed throughout the bubble, a significant fraction of the total volume contributes to the emission. Depending on the fraction $r$ (90, 95, 99\%) the emitting volume ranges from a few to almost 50\% of the spherical volume. We further illustrate the temporal evolution of the temperature in the bubble in Fig.~\ref{fig:pdf-temp-diff-times}, where we show the distribution of temperatures from the time of a prominent SN explosion at $t=10.93\,\mathrm{Myr}$ and the subsequent relatively long quiescent time up to $t=11.4\,\mathrm{Myr}$. We note the fast shift of the distribution from the explosion time to progressively lower temperatures with 0.5 dex colder values over a time span of only $500\,\mathrm{kyr}$.

We distinguish between times when a SN explodes, referred to as the \emph{active} state, and \emph{quiescent} states, during which the bubble evolves without recent SN impact. In Fig.~\ref{fig:overview-default-bubble}, we present three representative examples: an active state at time $t = 10.8\,\mathrm{Myr}$, and two quiescent states at $t=11.4\,\mathrm{Myr}$ and $t=12.5\,\mathrm{Myr}$. The second quiescent snapshot, chosen near the end of the simulation, corresponds to the longest time interval without internal SN activity.  In each set of panels, the first three rows show the density, temperature, and electron density in $x$--$y$ cuts through the center of the bubble. The fourth row displays the X-ray emissivity in the energy range $0.1-2\,\mathrm{keV}$. This highlights where the X-ray emission is produced within the bubble. Overlaid on the maps are contours that enclose the regions responsible for 80, 90 and 95 percent of the total X-ray emission. The fifth row shows line-of-sight projections of the X-ray emission. In case of a recent SN (left column, also showing the location of the last SN), we note that by far most of the emission originates from a very small volume around the explosion site. Contrary, during the quiescent states, the emission becomes more spatially extended. The earlier quiescent state ($t=11.4\,\mathrm{Myr}$) still retains residual structure from prior activity, while the later one ($t=12.5\,\mathrm{Myr}$) is characterized by the most volume-filling and diffuse emission, representing a more equilibrated post-SN phase.

\section{Impact of simplifications}
\label{sec:simplifications}

\begin{figure*}
    \centering
    \includegraphics[width=\textwidth]{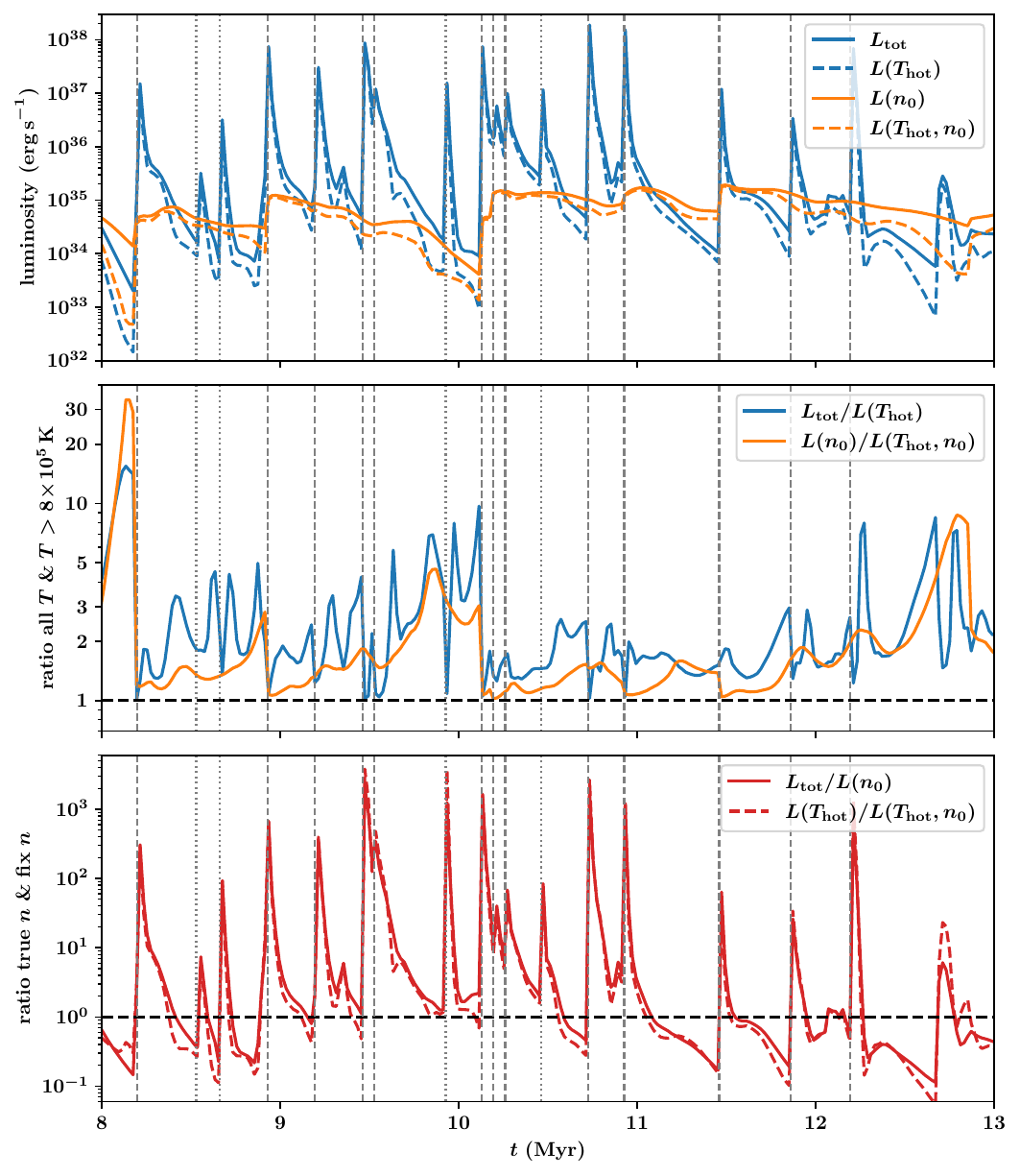}
    \caption{Comparison of X-ray luminosities under different modeling assumptions. The top panel shows the total luminosity evolution for various models (all gas with MHD properties, only hot gas, all gas at constant density, and hot gas at constant density). The middle panel presents the relative differences between these models, highlighting that during SN events, most of the emitting gas is at high temperatures, while in quiescent phases, contributions from cooler gas become significant. The bottom panel illustrates the impact of assuming a constant density, demonstrating that this simplification suppresses variability.}
    \label{fig:time-evol-lumi-assumtions}
\end{figure*}

We note that the X-ray luminosity increases by several orders of magnitude as soon as a SN goes off. This pronounced peak is of particular interest; however, it also introduces several challenges. On the one hand, numerical simplifications play a significant role. In our approach, we inject thermal energy into a predefined region selected as the site of a SN. This method does not account for the presence of massive stars prior to the SN event, nor does it include the effects of stellar winds or ionizing radiation. Consequently, the sudden heating of (partially) dense gas may lead to an overproduction of dense hot gas, potentially affecting the subsequent dynamical evolution and observable properties.

Beyond these numerical constraints, additional simplifications are necessary for the interpretation of the synthetic observables. A classical assumption in observational X-ray studies of bubbles is that the density within the bubble is uniform and that only gas with temperatures above $8 \times 10^5$ K contributes to the X-ray emission.

To assess the impact of these assumptions, we systematically compare different modeling approaches in Fig.~\ref{fig:time-evol-lumi-assumtions}, again analyzing a sphere with a radius of $100\,\mathrm{pc}$ and focusing on the energy range of $0.1-2\,\mathrm{keV}$. The top panel presents the luminosity evolution for various models, while the middle and bottom panels illustrate the relative differences between these models. We compute the total luminosity, $L_\mathrm{tot}$, by integrating over all gas cells, using the temperature, density, and ionization state directly obtained from the MHD simulations. In contrast, the model $L(T_\mathrm{hot})$ considers only gas with $T > T_\mathrm{hot} \equiv 8 \times 10^5$ K, while keeping the density, ionization degree, and temperature as in the simulations. To isolate the effects of a constant density assumption, we also compute models $L(n_0)$ and $L(T_\mathrm{hot},n_0)$, where in the former the density is fixed to $n_0 = 0.01\,\mathrm{cm}^{-3}$ in the entire sphere while keeping the temperature as is, and in the latter we fix the density to $n_0$ nd only consider hot gas.

We note that the experiments with constant density are intended to investigate the resulting overall differences. While it is true that in X-ray observations usually a constant density is assumed in order to infer the extent of the LB in a particular look direction, the observable remains the integrated emission from a line-of-sight within the X-ray band, containing gas of different densities and temperatures. As such, X-ray observations always measure $L_{\rm tot}$ or $L({T_{\rm hot}})$ rather than $L(n_0)$ or $L(T_{\rm hot}, n_0)$.

Several key trends emerge from this analysis. The most pronounced difference arises from the assumption of a fixed density, which significantly reduces the variability of the emission over time. In this case, peak luminosities are less pronounced, while low-emission phases tend to be overestimated. Although the absolute emission level depends on the chosen density value, the overall dynamical range is notably diminished when assuming a constant density. Moreover, restricting the emission to hot gas ($T > 8 \times 10^5$ K) primarily affects the quiescent phase, where overall emission levels are low. During these periods, warm gas slightly below the temperature threshold can contribute significantly to the total emission, an effect that is excluded in models that consider only the hot-phase gas.

The middle panel of Fig.~\ref{fig:time-evol-lumi-assumtions} quantifies these differences. The blue curves represent the ratio of $L_\mathrm{tot}$ to $L(T_\mathrm{hot})$ (solid vs. dashed blue lines in the top panel), while the orange curves show the ratio $L(n_0)/L(T_\mathrm{tot},n_0)$. Notably, the ratio $L_\mathrm{tot}/L(T_\mathrm{hot})$ approaches unity during SN events, indicating that, at these times, the majority of the gas is at temperatures exceeding $10^6$ K. In contrast, during quiescent phases, emission from cooler gas can dominate, leading to ratios between a few and ten. For the constant density models, the ratio $L(n_0)/L(T_\mathrm{tot},n_0)$ shows a smoother evolution. At the explosion times, the ratio does not jump to unity as often as in the other ratio and does not exceed values of $\sim3$. 

Finally, the bottom panel of Fig.~\ref{fig:time-evol-lumi-assumtions} compares the luminosities obtained with constant versus simulated density values. In general, models using the true simulated densities exhibit higher emission levels, except during the most quiescent periods. Additionally, we find that the average gas density within the bubble remains relatively stable over the simulation time, as shown in Fig.~\ref{fig:time_evol-therm-100pc}. This suggests that while density variations influence the emission properties, the assumption of a constant density may still provide reasonable estimates in certain conditions.

\section{All-sky maps of the simulated bubble}
\label{sec:skymaps}

We generate all-sky maps in Mollweide projections to illustrate various quantities as seen from an observer in the center of the bubble. To do so, we perform integrations along paths in different orientations of the sky and measure several relevant quantities. Before proceeding with the analysis and quantitative assessment, it is necessary to introduce some fundamental concepts and preliminaries.

\subsection{Column density effects}

Photoelectric absorption is the dominant process that attenuates X-rays in the energy range of interest; for a detailed review, see \citet{GorensteinTucker1976}. In general, the absorption follows a functional dependence that scales approximately with the $E^{-8/3}$, where $E$ is the energy. The energy-dependent cross section, $\sigma_e(E)$, per hydrogen atom was calculated by \cite{BrownGould1970} for a gas with solar composition, explicitly accounting for the contributions from heavier elements. A more detailed calculation of the cross section, particularly relevant for the lower-energy part of the soft X-ray band, was carried out by \citet{CruddaceEtAl1974}.

The impact of varying column densities of interstellar matter, $N_\mathrm{H}$, on an intrinsically simple power-law spectrum is significant. For $N_\mathrm{H} = 10^{20}\,\mathrm{cm^{-2}}$, the spectrum experiences attenuation of approximately one optical depth at $E = 0.2\,\mathrm{keV}$. Due to the $8/3$ power dependence, very little radiation is transmitted below $0.1\,\mathrm{keV}$, while absorption becomes negligible above $0.4\,\mathrm{keV}$. For $N_\mathrm{H} = 5\times10^{21}\,\mathrm{cm^{-2}}$, the interstellar medium effectively blocks all radiation below $0.1\,\mathrm{keV}$. Additionally, the transmitted spectrum exhibits a pronounced discontinuity at the absorption edge of interstellar oxygen.

We define the critical column density $N_\mathrm{crit}$ as the threshold above which most of the X-rays are absorbed. This value depends on details of the assumed spectrum as well as the chemical composition of the gas. To determine $N_\mathrm{crit}$ we use tabulated models based on \textsc{Xspec/tbabs} \citep{WilmAllenMcCray2000} (see Appendix~\ref{sec:Ncrit-tbabs}). For our temperature range we find values in the range $N_\mathrm{crit} = 10^{20}-10^{21}\,\mathrm{cm^{-2}}$. We chose $N_\mathrm{crit} = 10^{20}\,\mathrm{cm^{-2}}$ as our fiducial value, which sets a larger emphasis on the bubble interior. Tests with $N_\mathrm{crit} = 10^{21}\,\mathrm{cm^{-2}}$ only show minor differences, which we show and discuss in Section~\ref{sec:results}, Fig.~\ref{fig:luminosity-time-evol} and Appendix~\ref{sec:Ncrit-maps}. None of our conclusions depends on the exact number of $N_\mathrm{crit}$. We further emphasize that we do not account for an exact ray tracing from every X-ray emitting gas cell to all regions in the bubble of the simulation box. We simply use the critical column density to derive an approximate depth and the related effective X-ray properties.

\subsection{Emission measure}
\label{sec:EM}

The EM is a fundamental quantity in X-ray astronomy, describing the total X-ray emission from a hot, ionized plasma. It is defined as
\begin{equation}
\mathrm{EM} = \int n_e^2 \, \mathrm{d}V,
\end{equation}
where $n_e$ is the electron density and $\mathrm{d}V$ is the differential volume element. More generally, the EM reads $\int n_e n_{\rm H}dV$, accounting for the ions and electrons in a plasma. In our case the number density of free electrons corresponds to the one of protons, so the definitions are equivalent \citep[see also][]{Leahy2024}. Since X-ray emission depends on $n_e^2$, the EM provides a direct measure of the total emission strength from a given volume. Soft X-ray background (SXRB) studies using ROSAT \citep{1995ApJ...454..643S}, XMM-Newton \citep{Henley_2010_2}, and Chandra \citep{2003ApJ...583...70M} constrain the LB's EM to around $0.005 \, \mathrm{cm}^{-6} \, \mathrm{pc}$. 

\subsection{All-sky maps}

To generate all-sky maps we use \textsc{Healpix} \citep{GorskiEtAl2005,Zonca2019} to identify the position of each gas cell on the sky. In all cases we align the north pole of the all-sky map with the $z$ unit vector and let the $x$ unit vector point to $\phi=\theta=0$. In this orientation the disk midplane is located at $\theta\sim0$ (Galactic coordinates). The angular size of each cell is approximated via its linear extent ($\Delta x = V^{1/3}$) and the distance to the center of the bubble (observer). The cell is then mapped as a disk onto the plane of the sky. The luminosity at a given \textsc{Healpix} element $p_j$ on the sky is computed as
\begin{equation}
    L_X(p_j) = \sum_{\{i\}} l_i f_i,
\end{equation}
where $l_i = \varepsilon_i v_i$ is the contribution of the luminosity of the hydrodynamical cell $i$ given its emissivity $\varepsilon_i$ and volume $v_i$. The fraction $f_i$ encodes over how many \textsc{Healpix} elements $N(p_j)$, the hydrodynamical cell is spread, $f_i = 1/N_i(p_j)$. Finally, $\{i\}$ is the set of cells overlapping with the \textsc{healpix} cells at point $p_j$ along the line of sight. By including $f_i$ we ensure that the total luminosity $L_\mathrm{tot} = \sum_\mathrm{sphere} l_i = \sum_j L_X(p_j)$.  For the corresponding flux, we need to weigh the luminosity by the inverse square of the distance $d_i$ between the cell and the observer. Since cells close to the observer can have an artificially strong impact on the local flux and can also dominate large portions of the sky, we distinguish between \emph{close} cells ($d_i< d_\mathrm{min}$) and distant cells ($d_i\ge d_\mathrm{min}$), i.e.\ $F_X(p_j) = F_{X,\mathrm{close}} + F_{X,\mathrm{distant}}(p_j)$, where we set $d_\mathrm{min}=10\,\mathrm{pc}$, see Appendix~\ref{app:closest-cell}. All close cells are placed at a distance of $10\,\mathrm{pc}$ and their flux is uniformly distributed across the sky
\begin{equation}
    F_{X,\mathrm{close}} = \sum_{\{i\}_\mathrm{close}} \frac{l_i}{4\pi\,d_\mathrm{min}^2\,N_\mathrm{pix}},
\end{equation}
with $N_\mathrm{pix}=$ being the total number of \textsc{Healpix} elements and $\{i\}_\mathrm{close}$ is the set of all cells whose center is closer than $d_\mathrm{min}$ to the observer. For the distant flux, we take a similar approach as for the luminosity, 
\begin{equation}
    F_{X,\mathrm{distant}}(p_j) = \sum_{\{i\}_\mathrm{distant}} \frac{l_i}{4\pi\,d_i^2},
\end{equation}
with $\{i\}_\mathrm{distant}$ being the set of all distance hydrodynamical cells that overlap with the \textsc{Healpix} element $p_j$.

In addition to the luminosity and flux maps, we generate all-sky projections of the temperature structure as seen from an observer at the center of the bubble. Since the temperature varies along the line of sight, we compute a weighted mean temperature for each \textsc{Healpix} element \( p_j \) on the sky. The weighting can be performed in different ways depending on the physical relevance of the projected temperature.

A common choice is the column density-weighted temperature, which represents the average temperature along the line of sight weighted by the local gas density:
\begin{equation}
    T_{\mathrm{cd}}(p_j) = \frac{\sum_{\{i\}} T_i \rho_i \Delta x_i}{\sum_{\{i\}} \rho_i \Delta x_i},
\end{equation}
where $T_i$, $\rho_i$ and $\Delta x_i$ are the temperature, number density, and cell linear size of the hydrodynamical cell $i$, respectively.

Another physically motivated choice is the luminosity-weighted temperature, which emphasizes regions of strong emission:
\begin{equation}
    T_L(p_j) = \frac{\sum_{\{i\}} T_i l_i}{\sum_{\{i\}} l_i},
\end{equation}
where \( l_i \) is the luminosity contribution from cell \( i \). This weighting naturally highlights the thermodynamic state of the gas that dominates the observed emission.

For plasma diagnostics, we also compute the emission-measure-weighted temperature, which is particularly relevant for interpreting X-ray and H\(\alpha\) observations:
\begin{equation}
    T_{\mathrm{EM}}(p_j) = \frac{\sum_{\{i\}} T_i n_{\mathrm{e},i}^2 \Delta x_i}{\sum_{\{i\}} n_{\mathrm{e},i}^2 \Delta x_i}.
\end{equation}
This approach gives higher weight to dense, ionized regions that contribute most to the observed EM.

\subsection{X-ray travel distance}
\label{sec:bub-size}
\begin{figure}
    \centering
    \includegraphics[width=0.47\textwidth]{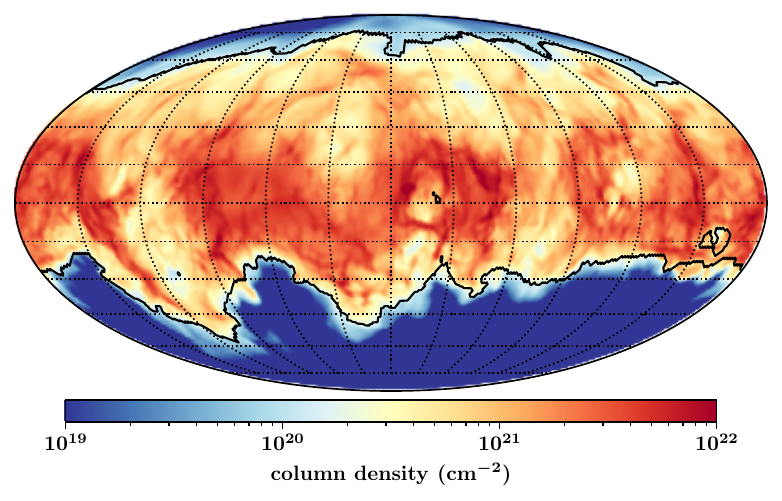}\\
    \includegraphics[width=0.47\textwidth]{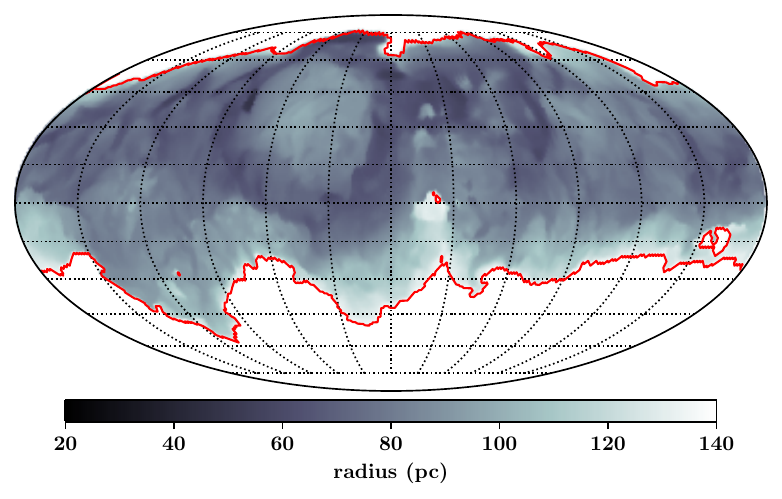}
    \caption{Illustration of the limited column density that X-rays can travel. In the top panel we show the total column density integrated from the center of the bubble out to a radius of $R_\mathrm{sp}=150\,\mathrm{pc}$. The contour lines represent the limit of a column density threshold of $N_\mathrm{crit}=10^{20}\,\mathrm{cm}^{-2}$. The bottom panel shows the corresponding radius out to which we need to integrate to reach $N_\mathrm{crit}$.}
    \label{fig:Mollweide-coldens-illustration}
\end{figure}

When computing the column densities from the center of the bubble to large distances, the integration through the dense shell surrounding the bubble exceeds the typical values of the maximum column density that soft X-rays can travel. We illustrate the effects in Fig.~\ref{fig:Mollweide-coldens-illustration} for a spherical region around the bubble center with a radius of $R_\mathrm{sp}=150\,\mathrm{pc}$, evaluated at our fiducial time of 10.8 Myr. Here we chose a larger radius as before because we would like to ensure that the bubble edges are included in this maximum radius. The top image shows the total column density integrated out to $R_\mathrm{sp}$, which exceeds the typical value of $N_\mathrm{crit}=10^{20}\,\mathrm{cm}^{-2}$. The threshold column density is shown in the black (red) contour line. We note that for the two caps the integration reaches the limit $R_\mathrm{sp}$ before $N_\mathrm{crit}$ is reached. The corresponding radius $R(N_\mathrm{crit})$ is shown in the bottom panel. From an observational perspective, the X-ray LB ends at this radius. The white regions in the radius plot indicate the open fraction of the bubble, i.e.\ where X-rays with $E\lesssim 0.1\,\mathrm{keV}$ from outside the bubble can reach the observer. 

\begin{figure}
    \centering
    \includegraphics[width=0.47\textwidth]{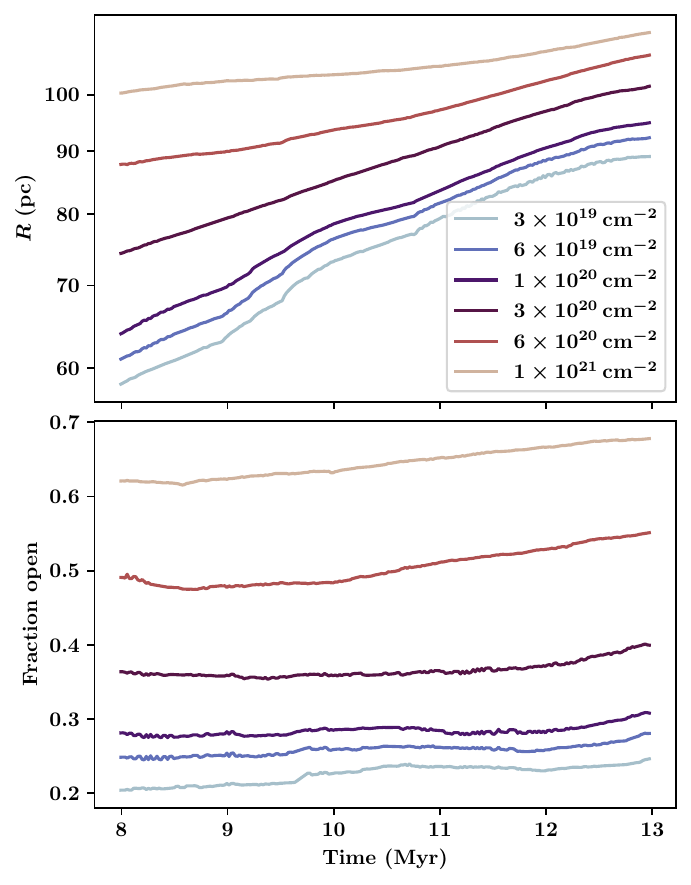}
    \caption{Time evolution of the median radius and the opening fraction of the shell for different column density thresholds $N_{\mathrm{crit}}$. The top panel shows the median distance from the center of the bubble to the location where the integrated column density along each line of sight reaches $N_{\mathrm{crit}}$. Higher column density thresholds correspond to larger median radii. The bottom panel shows the sky fraction below $N_{\mathrm{crit}}$ within $150\,\mathrm{pc}$. While the radius grows with time, the open fraction remains stable.}
    \label{fig:total-L-R}
\end{figure}

We extend the above illustration by taking into account different column densities, and show the evolution as a function of time. To do so, we determine the median radius for different column densities and different times, which is shown in Fig~\ref{fig:total-L-R}, top panel. As expected, the median radius increases with increasing column density, where a value of $N_\mathrm{crit}=3\times10^{19}\,\mathrm{cm^{-2}}$ yields median radii below $60\,\mathrm{pc}$ and a high column density of $10^{21}\,\mathrm{cm^{-2}}$ extends out to of $100\,\mathrm{pc}$ at the earliest time we consider. The radius increases by factor of approximately 1.5 over the investigated time span of five Myr. Choosing a critical column density of $10^{20}\,\mathrm{cm^{-2}}$ results in median radii from 75 up to $95\,\mathrm{pc}$. In the bottom panel, we show the fraction of the sky, for which the integration out to $150\,\mathrm{pc}$ does not reach the critical column density, i.e. the fraction of the bubble that is "open". This fraction depends again on $N_\mathrm{crit}$ with values as low as 0.2 for the lowest column density and values of 0.65 for the highest one. Contrary to the radius, which increases due to the overall expansion of the bubble the opening fraction of the sky remains stable over the simulated time.

\section{Results of the all-sky analysis}
\label{sec:results}

\begin{figure*}
    \centering
    \textbf{active state}\hspace{4.2cm}\textbf{quiescent state 1}\hspace{3.8cm}\textbf{quiescent state 2}
    \includegraphics[width=0.32\textwidth]{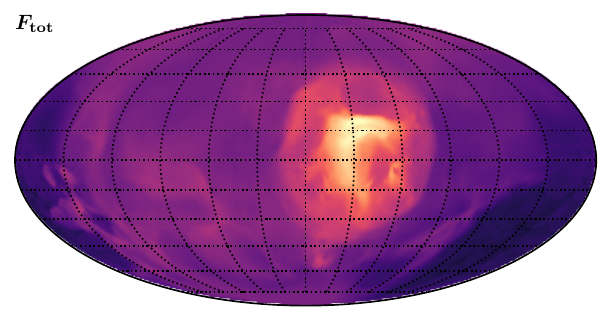}
    \includegraphics[width=0.32\textwidth]{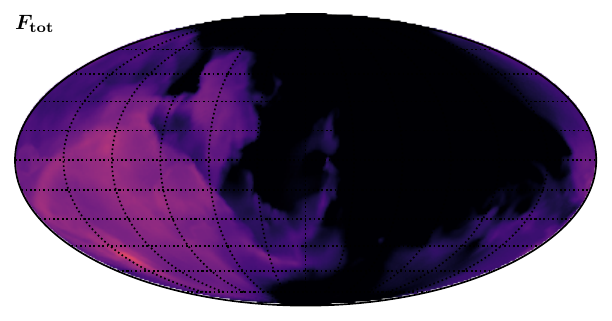}
    \includegraphics[width=0.32\textwidth]{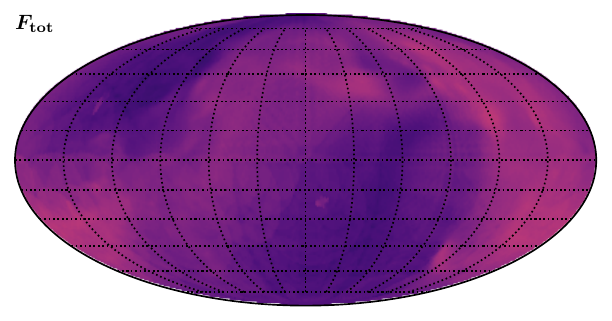}\\
    \includegraphics[width=0.32\textwidth]{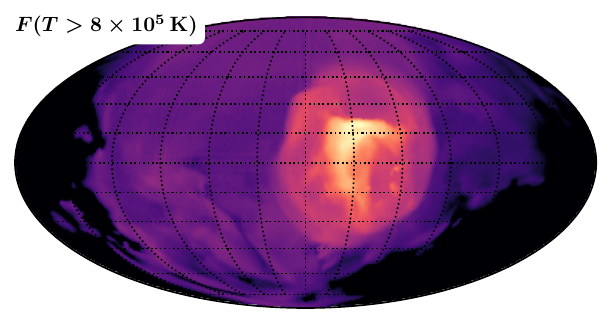}
    \includegraphics[width=0.32\textwidth]{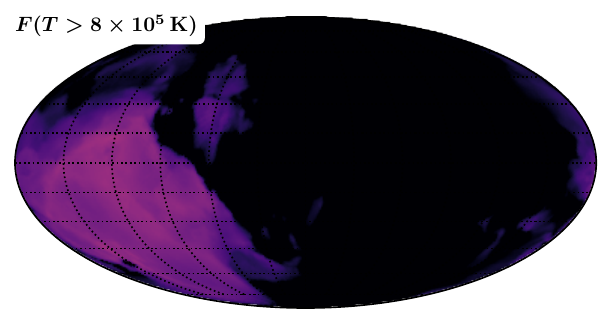}
    \includegraphics[width=0.32\textwidth]{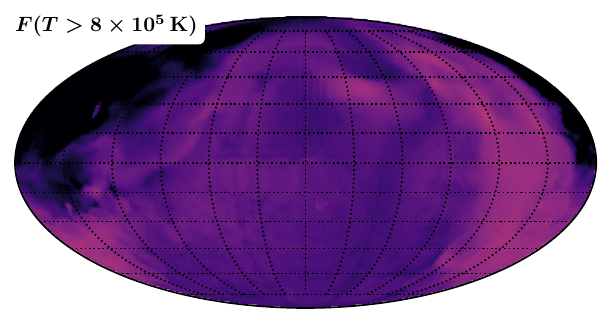}\\
    \includegraphics[width=0.32\textwidth]{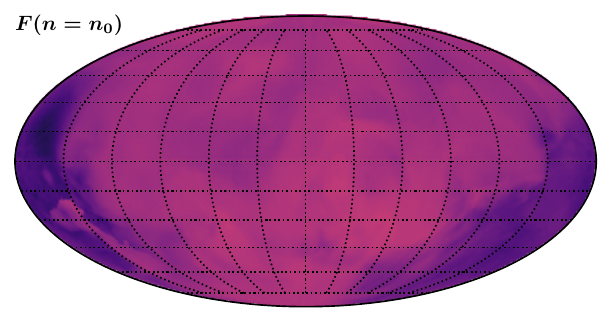}
    \includegraphics[width=0.32\textwidth]{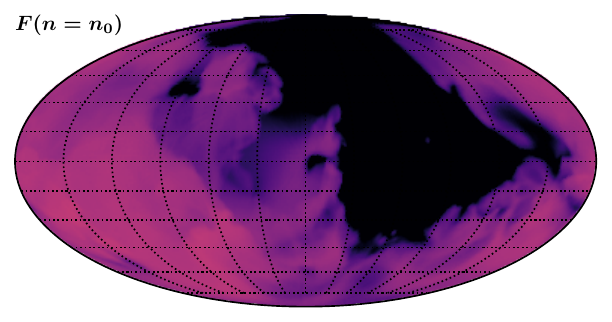}
    \includegraphics[width=0.32\textwidth]{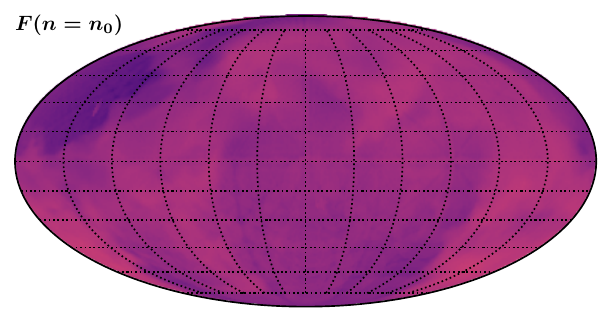}\\
    \includegraphics[width=0.32\textwidth]{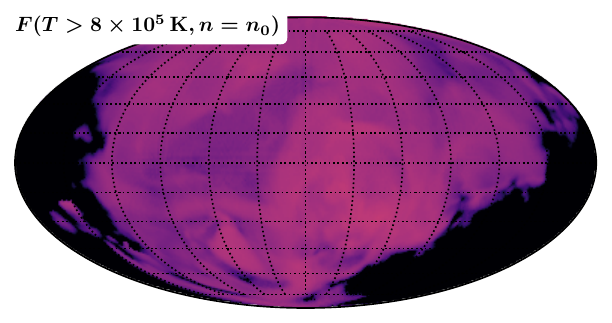}
    \includegraphics[width=0.32\textwidth]{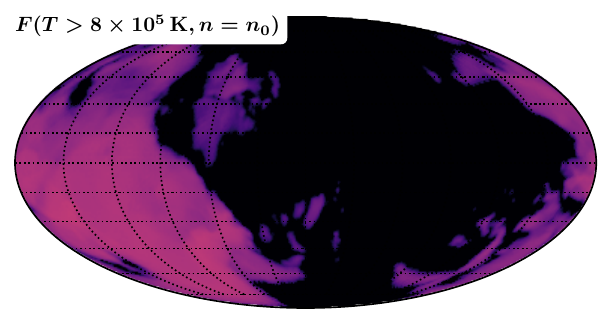}
    \includegraphics[width=0.32\textwidth]{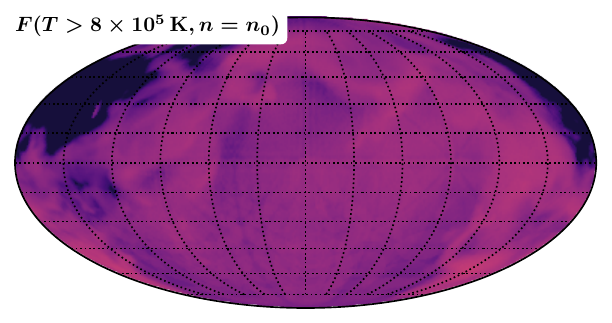}
    \includegraphics[width=0.9\textwidth]{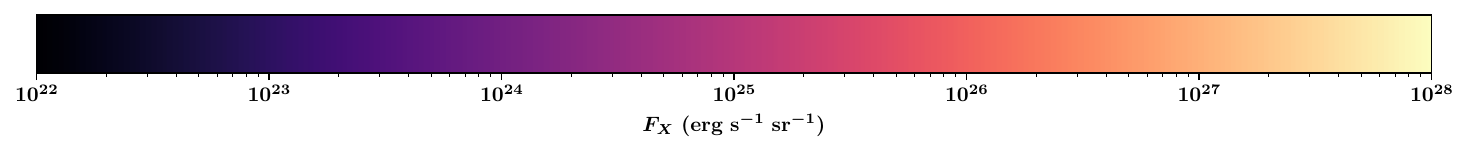}
    \caption{All-sky X-ray flux maps for the active state (left) and the two quiescent states (middle and right). In all cases we include the emission integrated up to the radius that corresponds to the threshold column density of $N_\mathrm{crit}=10^{20}\,\mathrm{cm}^{-2}$. From top to bottom we show the flux emitted from all gas, the flux coming from gas with $T>8\times10^5\,\mathrm{K}$, the flux from all temperatures but fixed number density ($n=n_0=0.01\,\mathrm{cm}^{-3}$), and the flux that originates only from hot gas ($T>8\times10^5\,\mathrm{K}$) assuming a uniform constant density of $n=n_0=0.01\,\mathrm{cm}^{-3}$. In the active state, the emission is strongly concentrated around the recent SN site. In the first quiescent state (middle column), the emission is fainter and more anisotropic, with several regions showing little or no flux. In contrast, the second quiescent state (right column) exhibits a more homogeneous and volume-filling emission pattern, indicative of a more relaxed configuration of the bubble interior.}
    \label{fig:flux-mollweide}
\end{figure*}

\begin{figure*}
    \centering
    \textbf{active state}\hspace{4.2cm}\textbf{quiescent state 1}\hspace{3.8cm}\textbf{quiescent state 2}
    \includegraphics[width=0.32\textwidth]{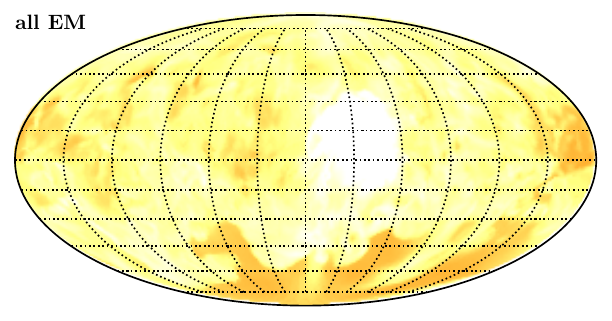}
    \includegraphics[width=0.32\textwidth]{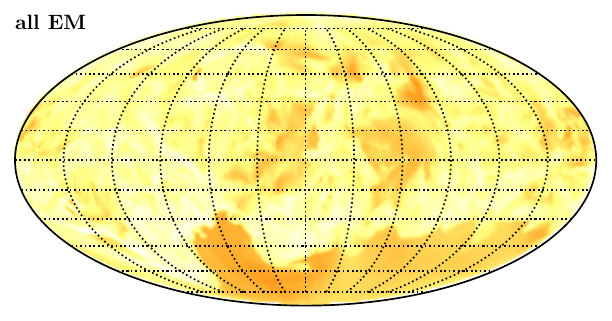}
    \includegraphics[width=0.32\textwidth]{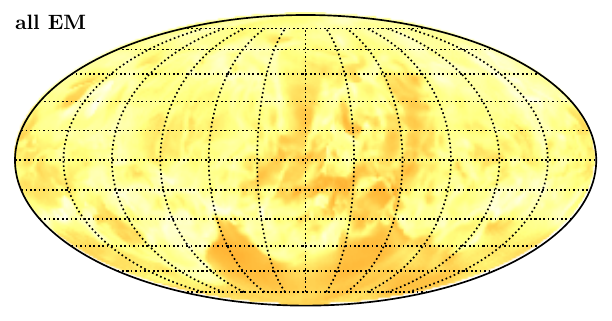}\\
    \includegraphics[width=0.32\textwidth]{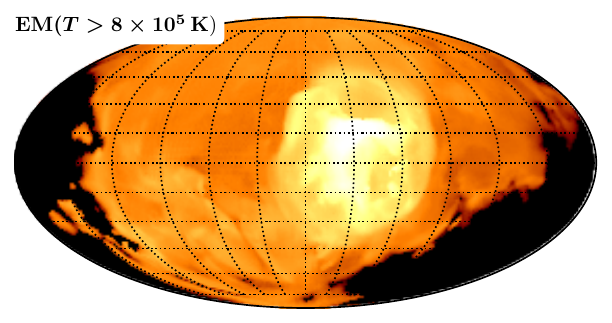}
    \includegraphics[width=0.32\textwidth]{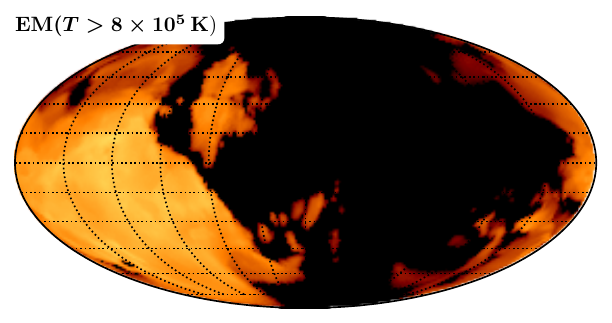}
    \includegraphics[width=0.32\textwidth]{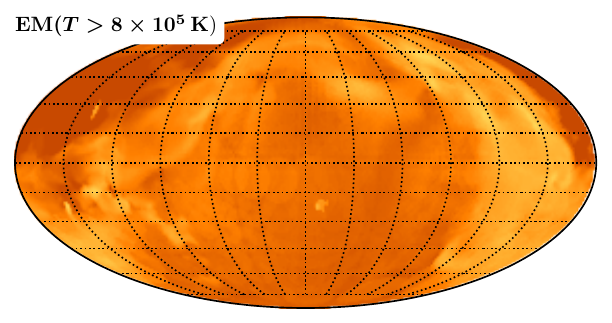}
    \includegraphics[width=0.9\textwidth]{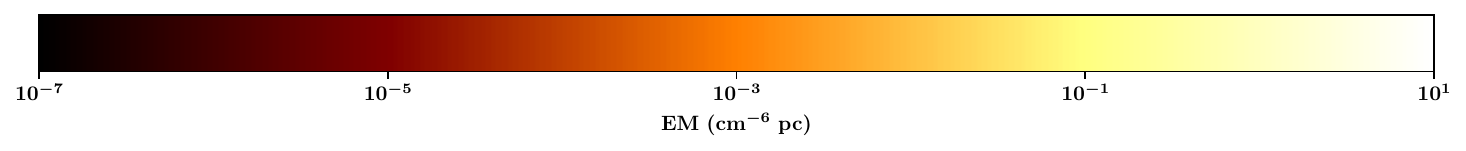}
    \caption{All-sky maps of the EM for the three different evolutionary stages. Again, the EM is integrated up to the radius corresponding to a column density threshold of $N_\mathrm{crit}=10^{20}\,\mathrm{cm}^{-2}$. The top panels show the contributions from all gas, the bottom ones only take the hot gas with $T>8\times10^5\,\mathrm{K}$ into account. Significant differences appear between the two rows, as the total EM is much higher and more widespread, while the EM from hot gas is more localized in the active state and more fragmented or absent in the quiescent phases. This is related to the temperature associated with the EM, see Fig.~\ref{fig:temp-mollweide}.}
    \label{fig:EM-mollweide}
\end{figure*}

\begin{figure*}
    \centering
    \textbf{active state}\hspace{4.2cm}\textbf{quiescent state 1}\hspace{3.8cm}\textbf{quiescent state 2}
    \includegraphics[width=0.32\textwidth]{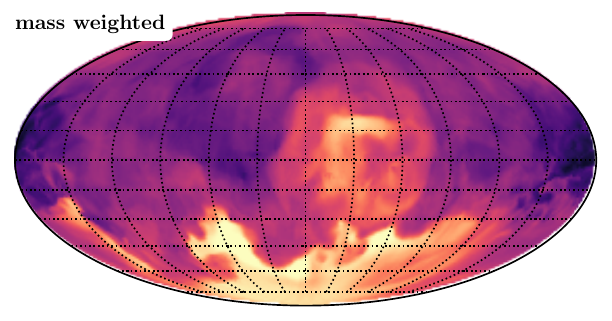}
    \includegraphics[width=0.32\textwidth]{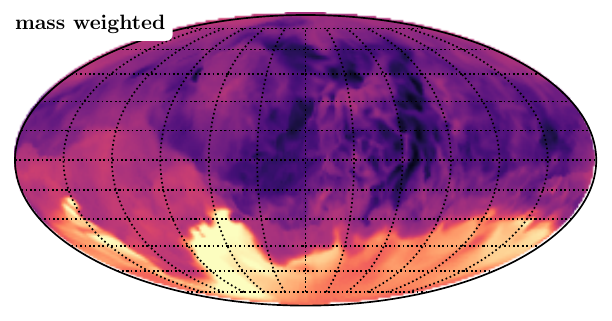}
    \includegraphics[width=0.32\textwidth]{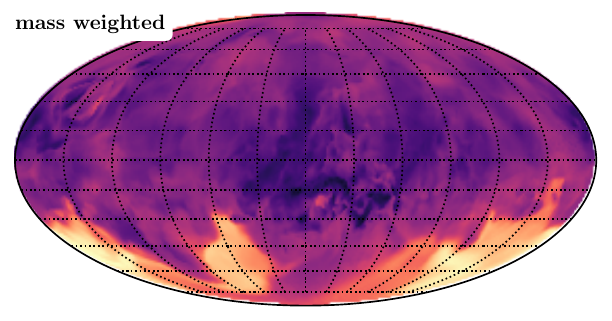}\\
    \includegraphics[width=0.32\textwidth]{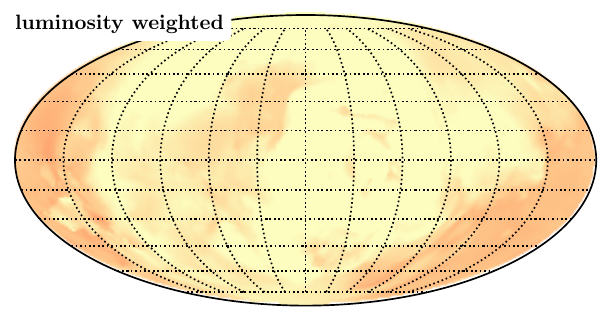}
    \includegraphics[width=0.32\textwidth]{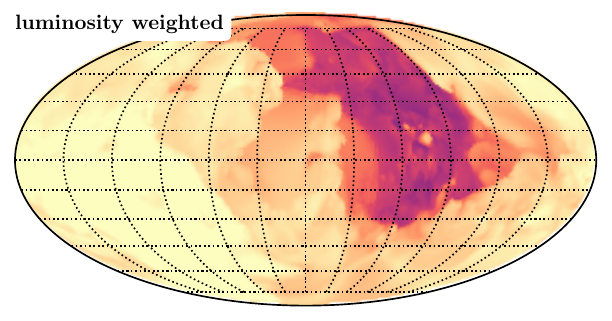}
    \includegraphics[width=0.32\textwidth]{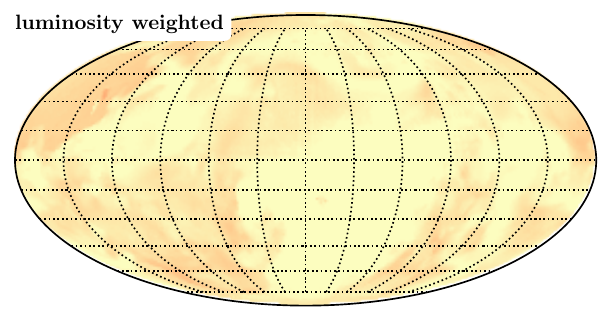}\\
    \includegraphics[width=0.32\textwidth]{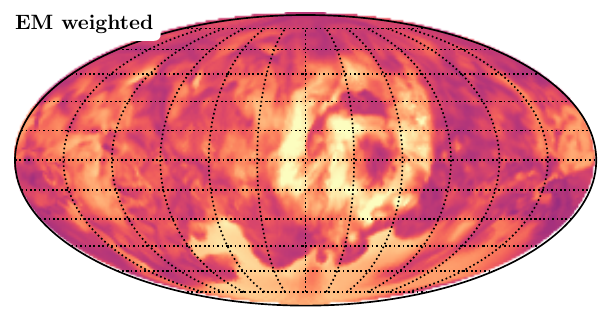}
    \includegraphics[width=0.32\textwidth]{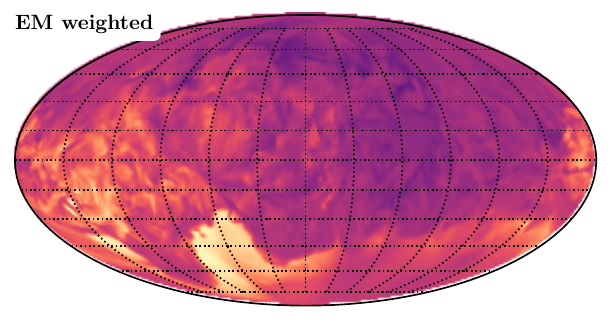}
    \includegraphics[width=0.32\textwidth]{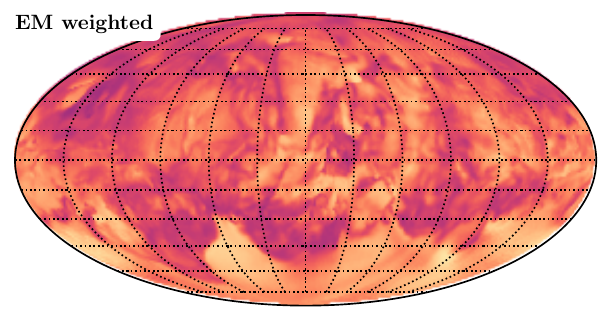}
    \includegraphics[width=0.9\textwidth]{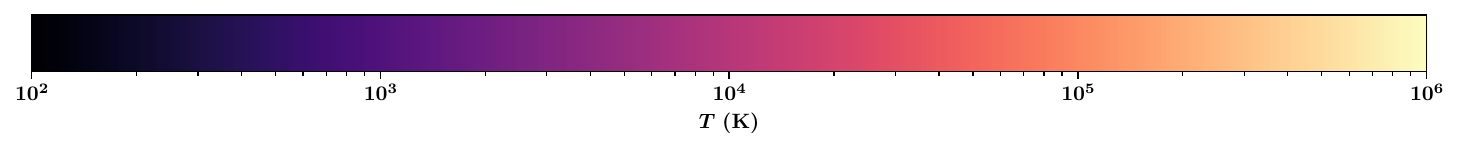}
    \caption{Temperature distribution across the sky, again integrated up to $N_\mathrm{crit}=10^{20}\,\mathrm{cm}^{-2}$. From top to bottom, the panels show line-of-sight temperature averages weighted by mass, X-ray luminosity, and M. The mass weighted temperatures are dominated by cool and warm gas ($T \sim 10^3$--$10^4$ K), with only modest variation between different states. The luminosity weighted temperatures are significantly higher ($T \sim 10^6$ K), reflecting the strong temperature dependence of X-ray emissivity. EM weighted temperatures lie in between, with values around $10^5$ K in the active state and lower, more spatially uniform temperatures in the quiescent states.}
    \label{fig:temp-mollweide}
\end{figure*}

\subsection{Comparison of active and quiescent state}

For the all-sky analysis we fix the column density to $N_\mathrm{crit}=10^{20}\,\mathrm{cm}^{-2}\approx1.67\times10^{-4}\,\mathrm{g\,cm}^{-2}$We keep the location of the observer fixed for all three analysis time but ensure that variations due to the picked center of the bubble do not influence our results (see Appendix~\ref{sec:fiducial-bubble-center}). We start by exploring the X-ray flux in Fig.~\ref{fig:flux-mollweide}, which shows all-sky maps for three different evolutionary stages: the active state (left column) and the two quiescent states (middle and right columns). Each column displays four different flux models: the total X-ray flux, the flux from hot gas with $T > 8 \times 10^5$ K, the flux assuming a constant density $n = n_0 = 0.01$ cm$^{-3}$ and finally the flux from hot gas under the same fixed density assumption. In the active state, the recent SN is visible in the total flux and the flux originating from hot gas (top two maps). It is not visible when fixing the number density to a uniform value. This highlights that the emission mainly originates from dense ($n>0.01\,\mathrm{cm}^{-3}$) heated gas. The region of the bubble opposite to the SN is relatively cold, which manifests in zero flux when only taking the hot gas into account (compare first and second panel on the left). In the first quiescent state (middle column), the emission is fainter and more anisotropic, with large regions of the sky showing no contribution from hot gas. This reflects residual inhomogeneities in temperature and density after the last SN events. By contrast, the second quiescent state (right column), taken later in the simulation, shows a more uniform and volume-filling emission pattern, especially in the total flux and the simplified models with constant density. This indicates a more relaxed configuration, where the remaining X-ray emission is spread more evenly across the sky. The region of the SN in the active state corresponds to the non-emitting region in the quiescent state~1 (see also Fig.~\ref{fig:overview-default-bubble}). This is true for all flux models independent of density and temperature simplifications. We note that fixing the density has the strongest impact for the overall emission. This is not surprising since in quiescent states the emission is dominated by the diffuse gas and originates from a larger volume of the bubble (see Fig.~\ref{fig:time_evol-therm-100pc}). A change in the density for the entire volume thus has a significant effect.

We further investigate the origin of the X-rays by analyzing the EM in Fig.~\ref{fig:EM-mollweide}, with the same order of columns as in Fig.~\ref{fig:flux-mollweide}. The top row displays the total EM computed from all gas, while the bottom row shows the EM considering only hot gas with $T > 8 \times 10^5$ K. The most prominent difference is the large discrepancy in absolute values between the two rows. When all gas is included, the EM reaches values of $\sim 0.1$ cm$^{-6}$ pc or higher, which is two orders of magnitude above typical observational estimates \citep{2017ApJ...834...33L}. This overestimate arises because much of the denser, cooler gas contributes to the EM, but not to the X-ray flux. If only considering the hot gas, the values drop significantly, showing values of order $10^{-3}\,\emunit$, in much better agreement with the observations. In the active state, hot gas is clearly present near the SN site, producing strong localized EM. In the first quiescent state, the sky is more irregular, with extended regions where no hot gas is present and the EM effectively vanishes. The second quiescent state, occurring later in the simulation, exhibits a more uniform and volume-filling hot component, resulting in a smoother EM distribution at observable levels.

There is another effect to consider. The analysis of the local density \cite[e.g.][see also middle panel of Fig.~\ref{fig:time_evol-therm-100pc}]{LinskyRedfield2021, ONeillEtAl2024} suggest that a large fraction of the gas mass is at $ 10^4$--$10^5\,\mathrm{K}$ ($\sim10^{-3}-10^{-2}\,\mathrm{keV}$), which is too cool to efficiently emit X-rays. This explains why the simulated EM ($\sim 0.1\,\emunit$) is much higher than the observed values ($(1-5)\times10^{-3} \,\emunit$) \citep{SnowdenEtAl2014,YeungEtAl2023}. The crucial aspect is that denser gas at moderate temperatures dominates the EM, but this gas is not visible in X-rays. This is consistent with the temperature distribution across the sky weighted by EM (see Fig.~\ref{fig:temp-mollweide} below), which shows that most of the EM comes from gas at $10^{4}$--$10^{5}$\,K, while the X-ray emission is primarily produced by hotter gas (middle panel). The discrepancy suggests that the EM alone does not fully trace the hot X-ray-emitting plasma but also includes contributions from lower-temperature gas that does not significantly affect the observed X-ray flux.

Finally,  we complete the comparison by analyzing the sky-projected temperature maps in Fig.~\ref{fig:temp-mollweide}, shown for the three evolutionary stages and under three different weighting schemes: mass-weighted (top row), luminosity-weighted (middle), and EM-weighted (bottom). As expected, when weighted by density the average temperature is cool/warm with values of order $10^3-10^4\,\mathrm{K}$. The differences between the active and quiescent states are relatively modest in this weighting, though slightly more structure appears during the active phase. In all cases, the south polar region features hotter temperatures, reflecting the open geometry of the bubble along that direction, where hot gas extends up to the integration limit of $150\,\mathrm{pc}$, which explains the constant values for the luminosity weighted temperature in this part of the sky. The luminosity weighted temperature is in almost all regions at the million degree level. This is not surprising since the emission strongly scales with the temperature and the dynamic range for the emission is large (see Fig.~\ref{fig:2D-histogram}). The EM-weighted temperatures fall between the previous two. During the active state, the temperature reaches $\sim10^5\,\mathrm{K}$ over much of the sky, with a distinct high-temperature feature near the recent SN location. In the first quiescent state, the temperatures drop and show a more fragmented pattern. By contrast, the second quiescent state exhibits smoother and more volume-filling EM-weighted temperatures, typically around $10^4$--$10^5,\mathrm{K}$. These values are critical for understanding the discrepancy between the total EM (which includes cooler, denser gas in the shell) and the EM derived from hot gas alone: much of the contribution to the total EM comes from gas at temperatures below the X-ray emission threshold of $8 \times 10^5,\mathrm{K}$.

\begin{figure*}
    \centering
    \includegraphics[width=\textwidth]{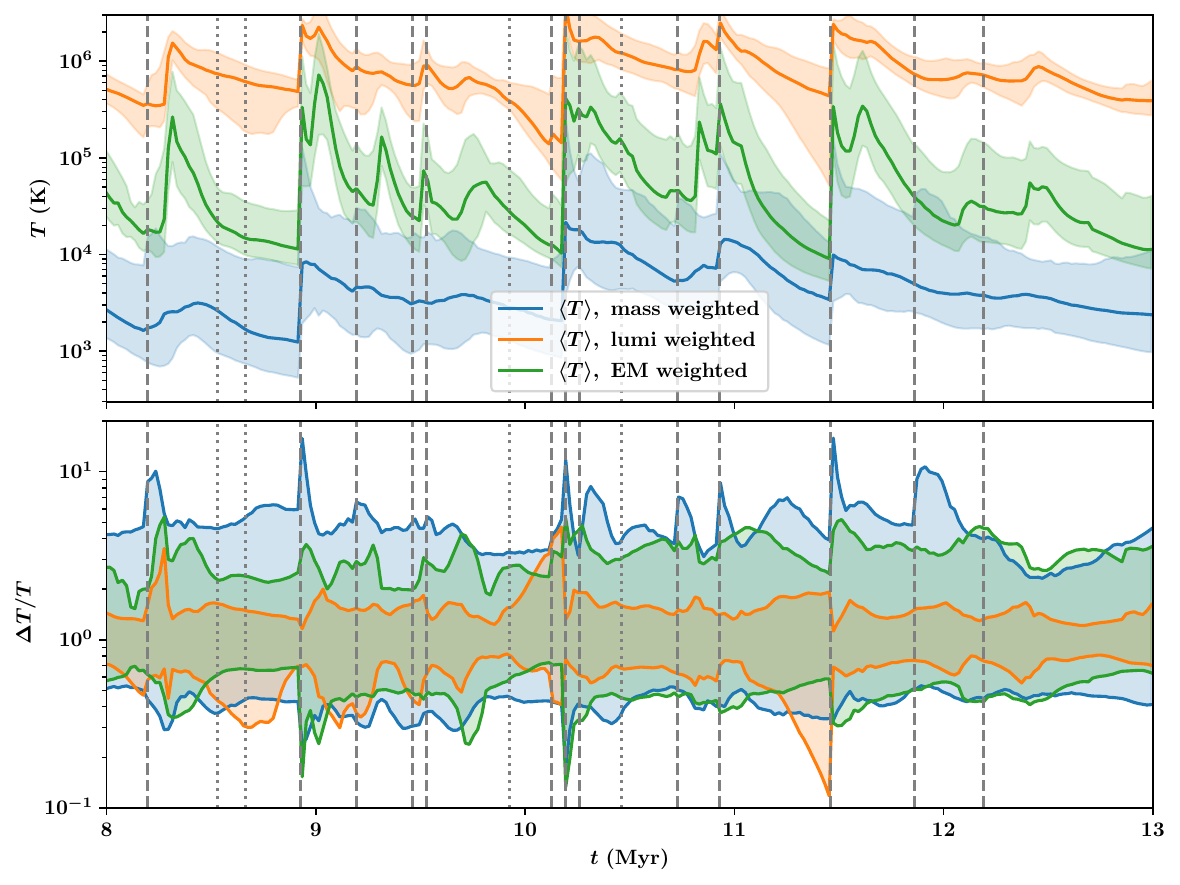}
    \caption{Time evolution of the temperature anisotropy on the sky, computed from the Mollweide projections at the fiducial critical column density of $N_\mathrm{crit}=10^{20}\,\mathrm{cm^{-2}}$. The top panel shows the median temperature ($T_\mathrm{med}$) for mass, luminosity and EM weighting. The shaded area indicates the 20$^\mathrm{th}$ ($T_{20}$) and 80$^\mathrm{th}$ percentile ($T_{80}$) of the distribution across the sky. The bottom panel shows the normalized values as shaded regions between $T_{20}/T_\mathrm{med}$ and $T_{80}/T_\mathrm{med}$. The vertical lines indicate the SNe as in Fig.~\ref{fig:time_evol-therm-100pc}.}
    \label{fig:anisotropy}
\end{figure*}

We note that the maps show a significant anisotropy on the sky. In the case of the active state, this is not surprising, since the location of the recent SN dominates the emission and results in a peak of the temperature. In the case of the quiescent state, the anisotropy depends on the hydrodynamical evolution of the bubble, which suggests a more homogeneous distribution of hot and warm gas. The degree of anisotropy of the temperature distribution over time is shown in Fig.~\ref{fig:anisotropy}. The top panel shows the median temperature ($T_\mathrm{med}$) with different weightings together with the range indicated as the shaded region marking the 20 ($T_{20}$) and 80 ($T_{80}$) percentile bounds. The bottom panel shows the dynamic range $\Delta T/T$ as the shaded region between $T_{20}/T_\mathrm{med}$ and $T_{80}/T_\mathrm{med}$. We find anisotropies that range from one tenth of the median temperature up to 10 times the median value. The largest dynamic range often correlates with SN events.

\begin{figure}
    \centering
    \includegraphics[width=\linewidth]{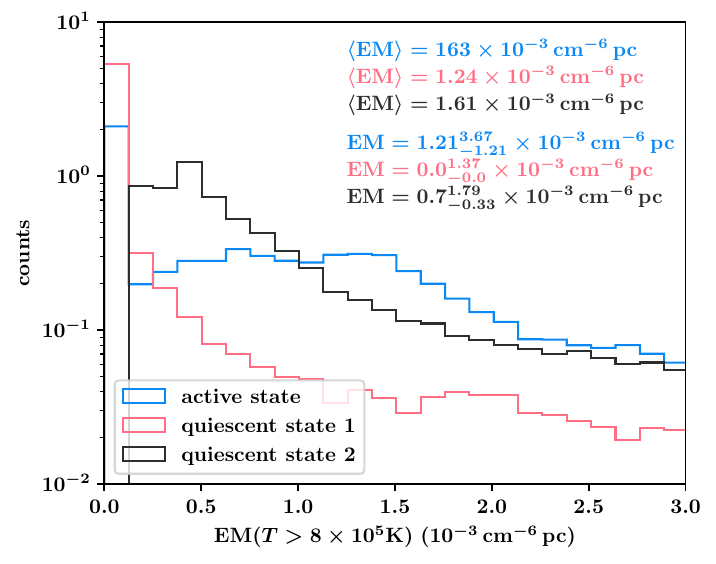}\\
    \includegraphics[width=\linewidth]{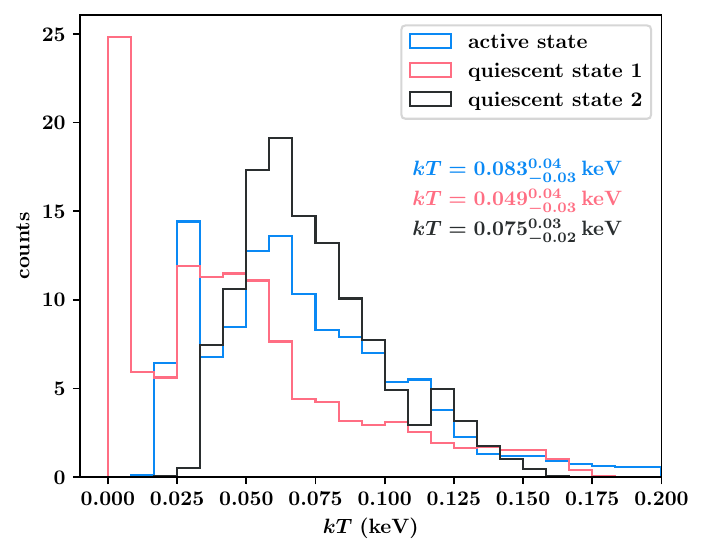}
    \caption{Histograms of the hot EM (top) and luminosity weighted temperature $kT$ (bottom) for the active state and the two quiescent states. For EM we mark the mean, as well as the median together with the 20$^\mathrm{th}$--80$^\mathrm{th}$ percentile range. For the temperature we show the median and its 20$^\mathrm{th}$--80$^\mathrm{th}$ percentile interval. The EM differs strongly between the active and the quiescent states: the active state has a mean EM of $163\times10^{-3}\,\mathrm{cm^{-6}\,pc}$, while the quiescent states exhibit values of $(1.2-1.6)\times10^{-3}\,\mathrm{cm}^{-6}\,\mathrm{pc}$. In contrast the temperatures differ much less. In the active state the median temperature is highest, but only by about 10--30 percent.}
    \label{fig:histograms}
\end{figure}

We compare the measured EM and temperature values to the recent observations by \citet{Yeung_2024}. We select again the three main snapshots in time and compute distributions of $\mathrm{EM}(T>8\times10^5\,\mathrm{K})$ and the luminosity weighted $kT$ in Fig.~\ref{fig:histograms}. The EM provides the clearest separation between the activity states: the active state shows a mean EM of $163\times10^{-3}\,\mathrm{cm^{-6}\,pc}$, whereas the two quiescent states reach only $(1.2-1.6)\times10^{-3}\,\mathrm{cm^{-6}\,pc}$, i.e.\ a contrast of two orders of magnitude. In contrast, the temperatures vary much less, with the active state exceeding the quiescent states by only $10$--$30\%$. These values differ from those reported by \citet{Yeung_2024}, who find EMs of $3.1\times10^{-3}$ and $3.5\times10^{-3}\,\mathrm{cm^{-6}\,pc}$ (a factor of a few higher than our quiescent values) and temperatures of $kT = 0.12$ and $0.14\,\mathrm{keV}$, which are higher than our results by roughly a factor of two.

\subsection{Time evolution}

\begin{figure*}
    \centering
    \includegraphics[width=\textwidth]{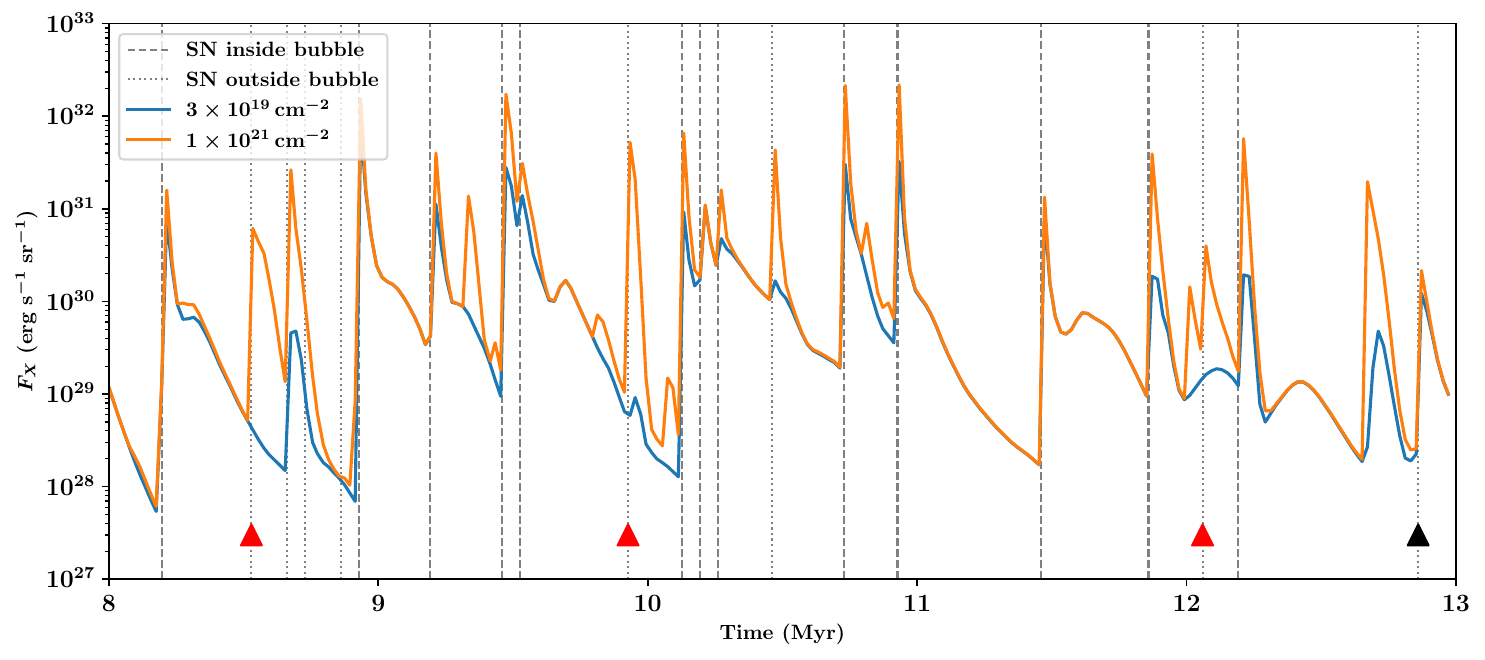}
    \caption{Time evolution of the total flux in the bubble. We highlight the evolution for two different extremes for the critical column density (orange vs. blue lines). The vertical lines indicate the exterior (dotted lines) and interior (dashed lines) SNe. The temporal variations are several orders of magnitude with peaks during SN explosions and a fast decline in the measured flux shortly thereafter. The difference between critical column densities depend on the location of the SNe and can be large if a SN explodes close to the bubble edge in regions of intermediate column density, see text. Some of the flux peaks might not be associated with the marked SN events, as they occur within 120 pc, but could instead result from more distant ($d > 120$ pc) SNe.}
    \label{fig:luminosity-time-evol}
\end{figure*}

In Fig.~\ref{fig:luminosity-time-evol} we revisit the time evolution of the total X-ray flux integrated up to two extreme values of the column density ($N_\mathrm{crit}=3\times10^{19}\,\mathrm{cm}^{-2}$, blue line, and $N_\mathrm{crit}=3\times10^{21}\,\mathrm{cm}^{-2}$, orange line). The vertical lines indicate again the SNe, where we distinguish between SNe that explode inside the bubble (dashed lines, $R_\mathrm{SN}\le100\,\mathrm{pc}$) and SN exploding outside in the shell surrounding the bubble (dotted lines, $100\,\mathrm{pc}<R_\mathrm{SN}<150\,\mathrm{pc}$). An important feature is the difference between SNe that explode inside the bubble versus the counterparts that explode outside it. The ones outside can still be seen in X-rays for the high column density threshold (orange lines). However, for the low column density threshold (blue lines) the shell of the bubble can be opaque to the emitted radiation depending on the location of the SN. The X-rays from external SNe that explode in regions with low integrated column density still reach the observer (e.g.\ the SNe at $t\sim12.9\,\mathrm{Myr}$, black triangle). Counter-examples are the ones at $t\sim8.5$, $9.9$, and $12.05\,\mathrm{Myr}$ (red triangles), where the blue curves continue to decline and the X-rays from those SNe are shadowed.

\section{Discussion and Caveats}
\label{sec:discussion}

\subsection{Comparison to observations}

The properties of our simulated analogue are overall consistent with measurements of the LB. 
The bubble's density ($n\sim10^{-2}\,\mathrm{cm}^{-3}$) and temperature ($T\sim10^5-10^6\,\mathrm{K}$, $\sim0.01-0.1\,\mathrm{keV}$) structure are in broad agreement with observational estimates, within a factor of 2--3 with observationally derived numbers \citep{LinskyRedfield2021, ONeillEtAl2024, Yeung_2024}. Given this basic agreement for the hot dilute medium, other derived properties of the Bubble are also in agreement. The closest distance from the observer to the edge of the bubble, defined as the length up to a column density of $N_\mathrm{crit} = 10^{20}\,\mathrm{cm}^{-2}$, is approximately 50\,pc, both in our numerical model and in \citet{Yeung_2024}. The distribution of distances in their analysis depends on the assumed dust data (\citealt{2022Lallement} vs. \citealt{Edenhofer2024}) but remains consistent with our distance map (Fig.~\ref{fig:Mollweide-coldens-illustration}). The sky covering fraction for $N_\mathrm{crit} = 10^{20}\,\mathrm{cm}^{-2}$ is approximately 30\%, again in agreement.

However, if we take the raw hydrodynamical simulation data, the derived EM is too high compared to observations. This discrepancy is particularly pronounced at the time of a recent SN event, when the (partially dense) gas in the shell is heated to strongly X-ray emitting temperatures. During this phase, the total EM reaches values as high as $10^3\,\mathrm{cm}^{-6}\,\mathrm{pc}$, significantly exceeding the observational range reported by \citet{Yeung_2024}, which lies between $1$--$5\times 10^{-3}\,\mathrm{cm}^{-6}\,\mathrm{pc}$. The discrepancy is less severe at a more quiescent time, when no recent SN has occurred, and the conditions resemble those of the actual LB more closely. However, even in this state, the EM remains too high, with a median value of $0.15\,\mathrm{cm}^{-6}\,\mathrm{pc}$ and a minimum around $0.1\,\mathrm{cm}^{-6}\,\mathrm{pc}$. The reason for this overestimation is that the EM calculation based on raw simulation data and includes all ionized gas, not just the hot gas responsible for X-ray emission ($T > 8 \times 10^5$ K). If we restrict the EM calculation to only the hot gas component, the derived values align well with the observed ones, resolving the discrepancy.

\subsection{Implications of our analysis}

Our results also have implications for observational studies of the LB. The large temporal variability of the soft X-ray luminosity, which spans almost four orders of magnitude between the quiescent and active phases, implies that the instantaneous luminosity is a poor tracer of the underlying physical conditions or formation history of the cavity. Instead of reflecting the integrated feedback over tens of Myr, the soft X-ray output is dominated by the most recent heating episode, such as a nearby SN or a transient phase of enhanced turbulent mixing. This sensitivity to short-lived events limits the ability of the observed luminosity to constrain either the initial conditions of bubble formation or the cumulative feedback processes that shaped the present-day structure.

This temporal behaviour suggests that the soft X-ray luminosity is better interpreted as a snapshot of the current activity level of the local environment rather than a fossil record of its past. Even modest changes in the energy injection rate or mixing efficiency lead to significant fluctuations in the X-ray output, whereas more robust structural quantities such as the cavity size, shell morphology, thermal pressure, and the distribution of warm ($10^4-10^5\,\mathrm{K}$) interface gas evolve more slowly and therefore provide more reliable tracers of the underlying feedback history. A meaningful comparison between simulations and observations should therefore focus on these slowly varying, multi-wavelength diagnostics (e.g.\ \textsc{O\,vi}, \textsc{N\,v}, dust extinction geometry, or the 3D distribution of neutral and molecular gas) rather than on the absolute level of soft X-ray emission alone.

\subsection{Caveats}

We note that the simulations only include SN feedback. As a result, early stellar feedback such as stellar winds and radiation cannot shape the bubble prior to the first SN. For the overall evolution of the bubble, this is only a minor drawback since the bubble is created around a massive star cluster, i.e.\ multiple SNe exploding in close proximity. The first few SNe therefore explode in an artificial environment, which is warm neutral gas rather than ionized gas by radiation and stellar winds. However, subsequent SNe explode in the hot dilute bubble, which in the end will determine the overall expansion of the bubble to the observed size. Concerning the total energetics, the size of the bubble and the shape, we do not expect the missing physics to have a significant impact.

The impact of the missing radiation from massive stars in our simulations can be assessed in this context. In low-density gas exposed to UV/EUV radiation from OB associations, photoionization equilibrium typically results in temperatures of order a few$\times 10^3-10^4\,\mathrm{K}$ in classical \textsc{H\,ii} regions and in the warm ionized medium, with somewhat higher equilibrium temperatures ($\sim 10^5\,\mathrm{K}$) possible in very dilute gas irradiated by a hard spectrum. Photoionization alone, however, does not usually maintain gas at $T \gtrsim 8\times10^5\,\mathrm{K}$; such temperatures are more naturally associated with shock heating and subsequent (non-)equilibrium cooling in a SN-driven cavity. The primary effect of including massive-star radiation would therefore be to sustain extended layers of warm and intermediate-temperature ($10^4$--$10^5\,\mathrm{K}$) gas and to modify the cooling history of shock-heated material, rather than to independently create a dominant $> 10^6\,\mathrm{K}$ phase.

Consequently, our current models likely underestimate the amount and longevity of intermediate-temperature, photoionized gas and may slightly underpredict the EM at the low-energy end of the soft X-ray band. They should, however, provide a reasonable lower bound on the genuinely hot ($T \gtrsim 10^6~\mathrm{K}$) component, which is primarily powered by SNe and shock heating. Incorporating radiation from massive stars in future radiation-MHD calculations will be essential to quantify how much additional hot gas can be maintained indirectly through altered cooling and mixing, and to refine the comparison between simulated temperature distributions, eROSITA-derived LB temperatures, and the observed soft X-ray spectrum of the LB.

Another limitation is that we only have one bubble in the simulation that matches the properties of the LB. Other bubbles driven by fewer SNe or different shapes are thus not included. Concerning the variability of the measured quantities, we compensate for the missing alternative bubbles by investigating the bubble over a total time of $5\,\mathrm{Myr}$, which includes 21 SNe exploding at different positions in the bubble as well as on the edges. The temporal variations thus allow us to mimic different bubble and SN configurations.

We do not include galactic shear in this isolated simulation box. For the long-term evolution of the magnetic field around the bubble and the long-term structure and its final dissolution we are therefore missing an important ingredient. Over the relatively short period of $5\,\mathrm{Myr}$ these effects are minor. The bubble is placed at the solar circle at a galacto-centric radius of $R_\mathrm{gal}\sim8\,\mathrm{kpc}$. The local circular velocity is $c_\mathrm{circ}\approx220\,\mathrm{km~s}^{-1}$ with a rather flat rotation curve \citep[e.g.][]{Jiao2023}. The time scale for shear to become important for a bubble with a size of $D=200\,\mathrm{pc}$ can be estimated vi $t_\mathrm{crit} = 2 \pi D / v_\mathrm{circ}\approx 5.6\,\mathrm{Myr}$. The evolution considered here is therefore at the upper limit, but still in the range to draw conclusions about the dynamics.

\section{Conclusions}
\label{sec:conclusions}

We analyzed a LB analogue from high-resolution MHD simulation of a multiphase ISM matching key observed properties of the LB. We performed post-processing of the soft X-ray emission, and investigating the bubble's temporal evolution as well as all-sky properties. Our findings can be summarizes as follows.

\begin{itemize}
    \item A key outcome of our simulation is the strong temporal and spatial variability of the X-ray emission. The X-ray luminosity in our LB analogue varies by several orders of magnitude on Myr timescales, strongly modulated by the occurrence of SN events. Immediately following a SN explosion, the X-ray luminosity can spike to values of order $L_X \sim 10^{38}\,\mathrm{erg\,s}^{-1}$, driven by the sudden heating of the surrounding gas. These bright phases are transient, typically decaying within $10^5$ years due to adiabatic cooling and the expansion of the remnant into lower-density regions. In contrast, quiescent phases are characterized by significantly reduced luminosities ($L_X \sim 10^{34}\,\mathrm{erg\,s}^{-1}$) and a more volume-filling distribution of the emitting gas.

    \item A critical outcome of our analysis is the realization that only a subset of the ionized plasma, specifically the high-temperature component ($T > 8 \times 10^5$ K), is responsible for the bulk of the X-ray emission. The broader thermal structure includes a significant amount of cooler, ionized gas that inflates the EM but contributes little to the actual radiative flux. This highlights the necessity of thermally selective approaches when connecting simulations to observations.
    
    \item The fraction of the bubble's volume responsible for the majority of the X-ray flux also evolves significantly with time. During active phases, approximately 90\% of the total X-ray flux is confined to only $\sim 1\%$ of the bubble volume, concentrated near the SN site. As the gas cools and disperses, the emission becomes more  spatially extended, with 30--40\% of the volume contributing to the same flux fraction during quiescent times.

    \item The detectability of the emission features is heavily modulated by column density effects along the line of sight. Due to photoelectric absorption, soft X-rays below $\sim 0.3$ keV are efficiently absorbed when the integrated hydrogen column density exceeds $N_\mathrm{H} \sim 10^{20}\,\mathrm{cm}^{-2}$. As a result, SN explosions that occur in regions surrounded by dense shells, either inside the bubble or just outside, can be entirely obscured from an observer located at the bubble center, depending on the local geometry and orientation. This leads to a strongly uneven distribution of observable X-ray emission across the sky, suggesting that many SN-driven X-ray peaks may go undetected depending on the observer's position and the structure of the surrounding ISM. However, while our simulations show significant temporal anisotropy in the X-ray emission, observations of the LB suggest a more isotropic emission morphology. This discrepancy might indicate missing physical processes in the model, such as stellar winds or non-equilibrium ionisation effects, which could lead to a smoother, more volume-filling distribution of hot gas,
    Synthetic observations created with different $N_\mathrm{crit}$ thresholds around the observer confirm that due to the local distribution of dense material, the same SN event may appear either prominently or be completely invisible in soft X-rays.
\end{itemize}

By tracing how transient heating and evolving density structures shape the observable emission, this work provides a dynamic perspective that complements static or time-averaged approaches. Future extensions of this work could incorporate additional physical processes, such as stellar winds or non-equilibrium Ionization effects, to support a more accurate comparison with current and future X-ray surveys.

\section*{Acknowledgements}
We thank the two anaonymous referees for a careful reading of the manuscript and detailed questions, which helped improving the paper noticeably. RSK and PG acknowledge financial support from the European Research Council via the ERC Synergy Grant ``ECOGAL'' (project ID 855130),  from the German Excellence Strategy via the Heidelberg Cluster of Excellence (EXC 2181 - 390900948) ``STRUCTURES'', and from the German Ministry for Economic Affairs and Climate Action in project ``MAINN'' (funding ID 50OO2206). The team in Heidelberg is grateful for computing resources provided by the Ministry of Science, Research and the Arts (MWK) of the State of Baden-W\"{u}rttemberg through bwHPC and the German Science Foundation (DFG) through grants INST 35/1134-1 FUGG and 35/1597-1 FUGG, and also for data storage at SDS@hd funded through grants INST 35/1314-1 FUGG and INST 35/1503-1 FUGG. RSK also thanks the 2024/25 Class of Radcliffe Fellows for their company and for highly interesting and stimulating discussions. RSK also thanks for computing resources provided by the Leibniz Rechenzentrum via grants pr32lo, pr73fi and GCS large-scale project 10391. MCHY acknowledge support from the Deutsche Forschungsgemeinschaft through the grant FR 1691/2-1. EM was co-funded by the European Union (ERC, ISM-FLOW, 101055318). The authors gratefully acknowledge the data storage service SDS@hd supported by the Ministry of Science, Research and the Arts Baden-Württemberg (MWK) and the German Research Foundation (DFG) through grant INST 35/1503-1 FUGG.

\section*{Data Availability}

The data is in parts available at the \href{http://silcc.mpa-garching.mpg.de}{SILCC data webpage, DR6}. Additional data files will be shared upon reasonable request to the corresponding authors.

\bibliographystyle{mnras}
\bibliography{references.bib}

\begin{appendix}

\section{SN positions}
\label{app:SN-positions}
\begin{figure}[h]
    \centering
    \includegraphics[width=8.4cm]{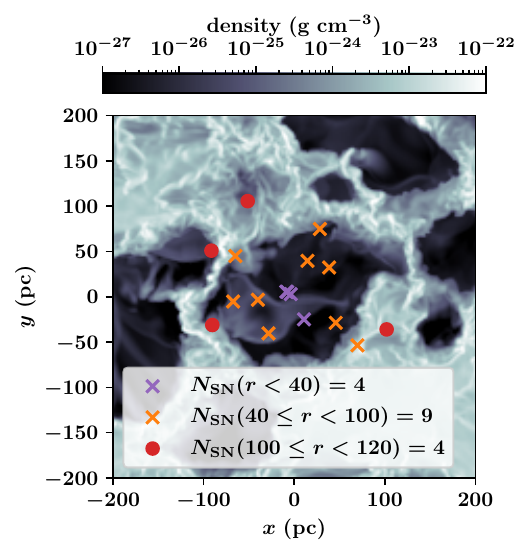}
    \caption{SN locations for our bubble over the time windows from $t=8-13\,\mathrm{Myr}$ split into different radii around the center of the bubble.}
    \label{fig:SN-positions}
\end{figure}

In order to better characterize the temporal and spatial structure of the X-ray emission, we analyze the distribution of SN explosion sites within and around the simulated bubble. In Fig.~\ref{fig:SN-positions} we show the projected positions of all SNe occurring between $t=8\,$Myr and $t=13\,$Myr, categorized by radial distance from the center of the bubble.
We distinguish between three radial zones: SNe within $r<40\,$pc (purple crosses), those between $40\le r<100\,$pc (orange crosses), and those in the shell region with $100\le r<120\,$pc (red circles). In total, the figure highlights 17 SNe contributing to the evolution of the bubble structure and the resulting X-ray emission.

\section{Comparison between \textsc{cloudy} and \textsc{apec} tables}
\label{sec:cloudy-vs-apec}

\begin{figure*}
    \centering
    \includegraphics[width=\textwidth]{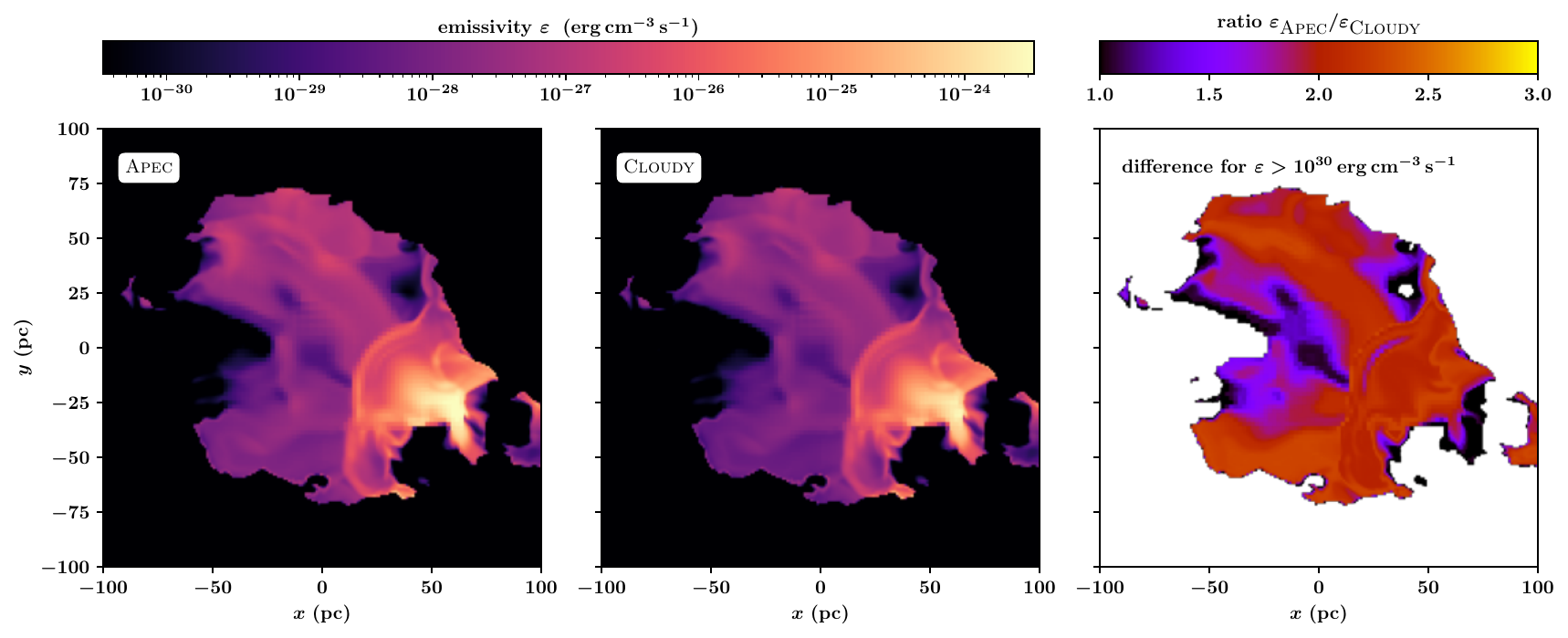}
    \caption{Comparison of the two different emissivity tables generated using \textsc{Apec} (left) and \textsc{Cloudy} (middle) as well as their ratio (right). Shown is the emissivity in a cut though the center of the bubble at the active state. The ratio is depicted in regions, for which the emissivity is above a threshold of $\varepsilon>10^{-30}\,\mathrm{erg\,cm^{-3}\,s^{-1}}$. At emissivities below this threshold, the relative difference can be large, but is irrelevant overall. The ratio shows that the \textsc{Apec} emissivities are approximately twice as high as the \textsc{Cloudy} counterparts. }
    \label{fig:comparison-tables-maps}
\end{figure*}

We compare two different emission tables in Fig.~\ref{fig:comparison-tables-maps} showing the \textsc{Apec} (left) and the \textsc{Cloudy} table (middle) as well as the ratio of both tables (right) for the snapshot of the active state. The variations between the maps are visually hard do distinguish. The \textsc{Apec} table shows higher emission with a ratio of $\varepsilon_\textsc{Apec}/\varepsilon_\textsc{Cloudy} \sim2$.

\section{Projected radial velocity}
\label{sec:mollweide-vrad}

\begin{figure*}
    \centering
    \textbf{active state}\hspace{4.2cm}\textbf{quiescent state 1}\hspace{3.8cm}\textbf{quiescent state 2}
    \includegraphics[width=0.32\textwidth]{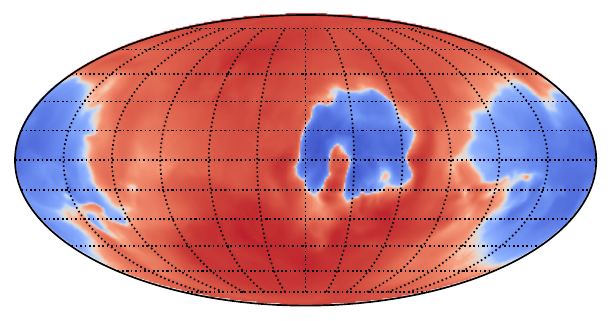}
    \includegraphics[width=0.32\textwidth]{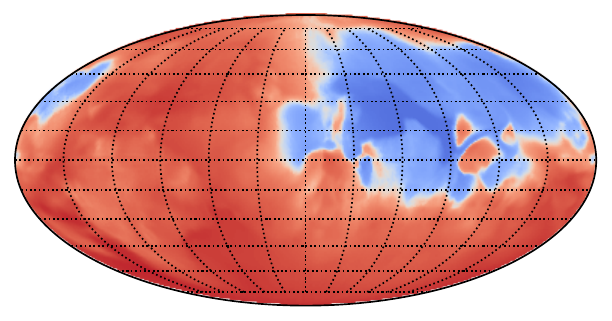}
    \includegraphics[width=0.32\textwidth]{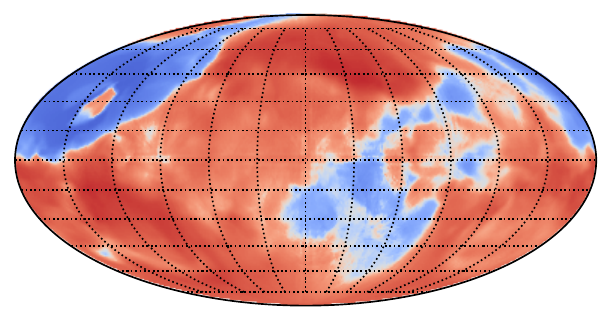}\\
    \includegraphics[width=0.32\textwidth]{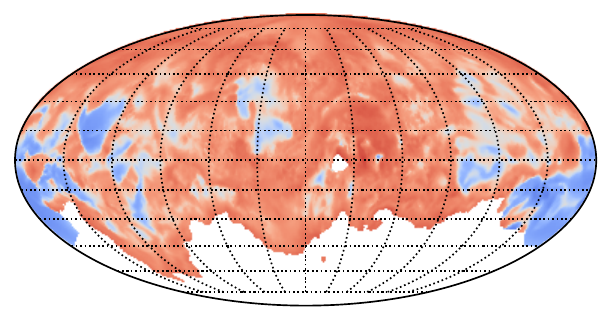}
    \includegraphics[width=0.32\textwidth]{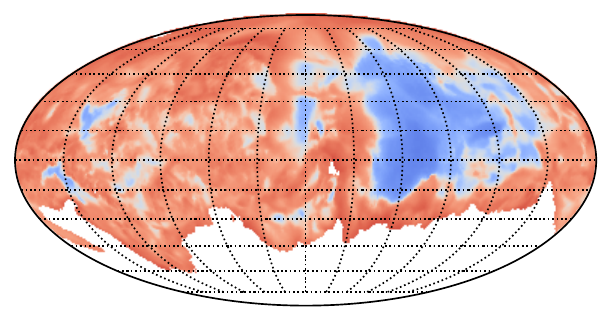}
    \includegraphics[width=0.32\textwidth]{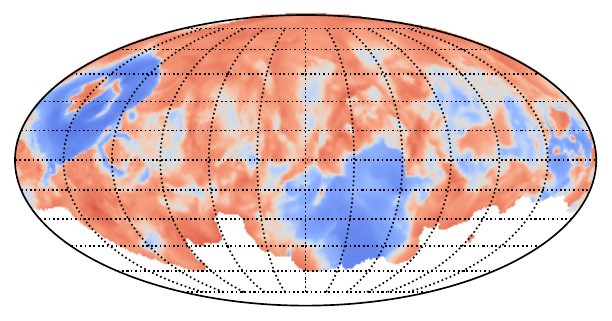}
    \includegraphics[width=0.9\textwidth]{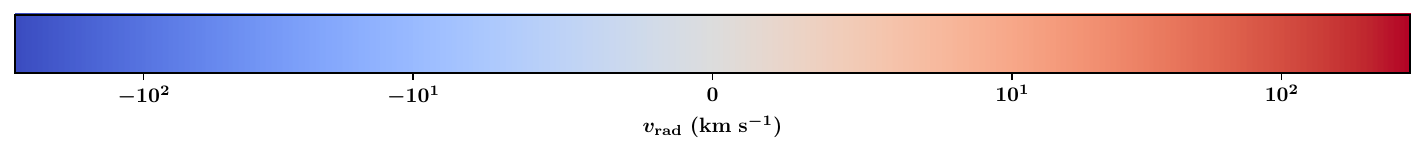}
    \caption{All-sky projections of the radial velocity with respect to the fiducial center of the bubble. From left to right we show the active and the two quiescent states. The top row shows the gas with an ionization fraction of $f_\mathrm{ion}>0.5$, i.e.\ the hot bubble gas. The bottom plots show the gas in the shell with ionization fraction between 0.01 and 0.5. The white areas in the south indicate regions, for which we do not reach the critical column density, i.e. we do not see a shell, cf. Figs.~\ref{fig:bubble-time-evol} and \ref{fig:Mollweide-coldens-illustration}. We note that the bubble gas is dominated by large patches of fast moving gas. In the active state the recent SN located at $\phi\sim\theta\sim30^\circ$ manifests in fast motions moving towards the observer (blue) without any counterpart in the shell. The two quiescent states show no direct recent SN activity, so the blue patches correspond to holes in the bubble wall through which gas from neighboring bubbles enters the main bubble.}
    \label{fig:mollweide-vrad}
\end{figure*}

In this section we illustrate the dynamics of the bubble in Mollweide projections of the radial velocity with respect to the center of the main bubble. The projections are shown in Fig.~\ref{fig:mollweide-vrad}. We again investigate the three main times, the active state at $t=10.8\,\mathrm{Myr}$ and the two quiescent states at $t=11.4$ and $t=12.5\,\mathrm{Myr}$. We compute the projected radial velocity in a sphere with radius $r=120\,\mathrm{pc}$ around the center of the bubble. Here, we do not apply a column density limit as in the main part of the paper, in order to actually show the full region around the center independent of the thickness of the shell and the accumulated gas. We distinguish between the hot gas with an ionization degree larger than $f_\mathrm{ion}=0.5$ (top row) and the more or less neutral gas ($0.5>f_\mathrm{ion}>0.01$, bottom row), where the former one represents the dilute gas and the latter shows the gas in the shell. We note that at later times, the bubble opens up and allows gas to enter the main bubble illustrated by the blue patches in the two quiescent states that spatially coincide in the top and bottom panels. The SN that just exploded in the active state pushes gas to the observer in the interior of the bubble (top left). This gas originates from within the bubble, so there is no open location in the bottom left shell structure (no corresponding blue spot).

\section{Location of the observer inside the bubble}
\label{sec:fiducial-bubble-center}

\begin{figure}
    \centering
    \includegraphics[width=\linewidth]{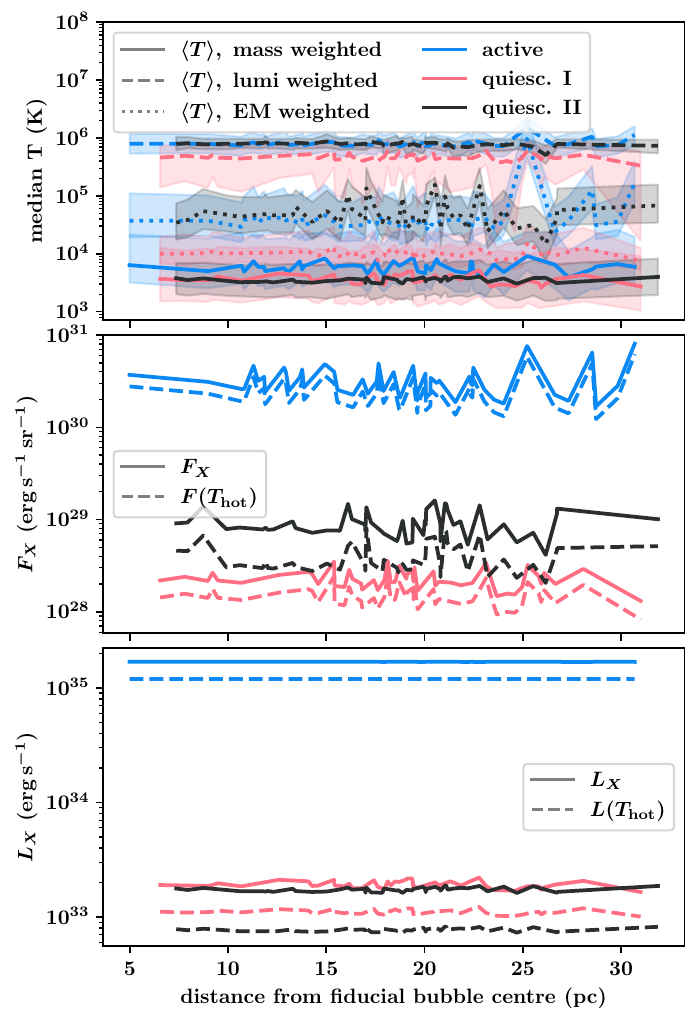}
    \caption{Impact of the observer's location on various diagnostic quantities as a function of distance. Results are shown for all three main analysis times. For each distance, we plot the median temperature over the Mollweide map together with the 25$^\mathrm{th}$--75$^\mathrm{th}$ percentile range (shaded), as well as the total integrated flux and the total luminosity across the sky. While these quantities vary by factors of a few with observer distance, we find no systematic trend. Because these spatial variations are smaller than the temporal variations between the three states, the precise observer position does not mask or dominate the temporal evolution.}
    \label{fig:location-variations}
\end{figure}

We picked the center of the bubble by eye and keep this point fixed over the entire evolution. To ensure that quantities sensitive to the distance between gas cells and the observer are not significantly affected by the choice of the fiducial center, we vary the assumed bubble center and recompute these quantities. For each of the three main analysis times (active state and the two quiescent states) we perform 50 additional Mollweide projections for a critical column density of $N_\mathrm{crit}=10^{20}\,\mathrm{cm}^{-2}$, where we vary the center by a random offset in $x$, $y$ and $z$ in the range from $-20$ to $20\,\mathrm{pc}$ around the fiducial center of the bubble. We show the median temperature as well as the total flux and the luminosity across the Mollweide map in Fig.~\ref{fig:location-variations} as a function of distance from the fiducial bubble center. There are fluctuations with distance, but the overall lines are flat indicating that the detailed location of the observer has a minor impact.

\section{Estimates for the critical column density}
\label{sec:Ncrit-tbabs}

\begin{figure}
    \centering
    \includegraphics[width=\linewidth]{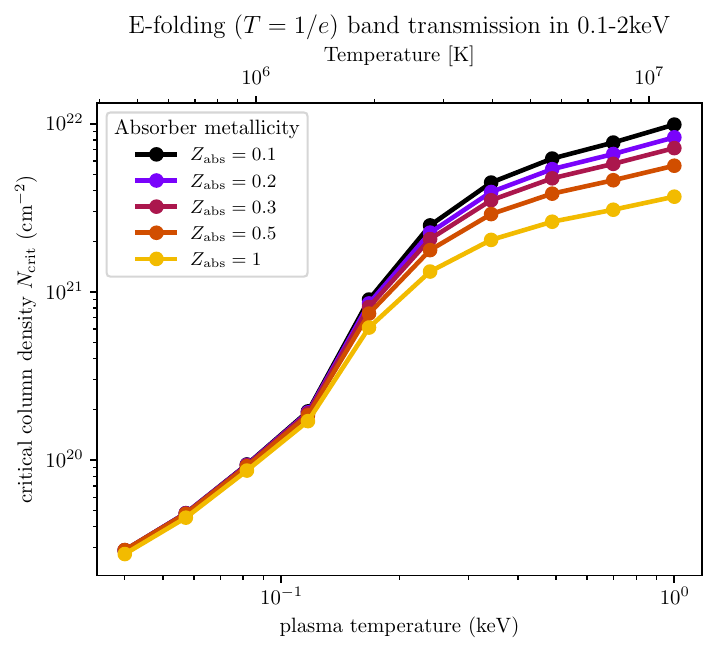}
    \caption{Critical column density as a function of plasma temperature using the \textsc{Apec} model in \textsc{Xspec/tbabs}. We show the cuves for several metallicities to account for uncertainties. For our temperatures we expect a critical column density in the range $N_\mathrm{crit} = 10^{20}-10^{21}\,\mathrm{cm}^{-2}$.}
    \label{fig:Ncrit-xspec-tbabs}
\end{figure}

We assume the critical column density to be $N_\mathrm{crit}=10^{20}\,\mathrm{cm}^{-2}$, over which a significant fraction of the X-rays are absorbed. Since the absorption properties vary depending on the spectral model, thermodynamic state of the gas as well as metallicity, we vary $N_\mathrm{crit}$ for some of the analysis in the main text. There are, however, uncertainties in the metallicity, which might be relevant for $N_\mathrm{crit}$. We compute $N_\mathrm{crit}$ corresponding to an e-folding value of the transmission
\begin{equation}
    N_\mathrm{crit} = N(T_0 \rightarrow T_0/e)
\end{equation}
as a function of plasma temperature for a range of metallicities, $Z_\mathrm{abs}\in [0.1:1]$ using the \textsc{tbabs} models \citep{WilmAllenMcCray2000} of the \textsc{Xspec} software package The resulting values for $N_\mathrm{crit}$ range from $3\times10^{19}-\times10^{22}\,\mathrm{cm}^{-2}$ and are shown in Fig.~\ref{fig:Ncrit-xspec-tbabs}. In the temperature range that is most relevant to our simulation data ($\sim10^6\,\mathrm{K}$), the critical column density is between $10^{20}$ and $10^{21}\,\mathrm{cm}^{-2}$ with only a minor impact of the assumed $Z_\mathrm{abs}$.

\section{Impact of the critical column density}
\label{sec:Ncrit-maps}

\begin{figure*}
    \centering
    \textbf{active state}\hspace{4.2cm}\textbf{quiescent state 1}\hspace{3.8cm}\textbf{quiescent state 2}
    \includegraphics[width=0.32\textwidth]{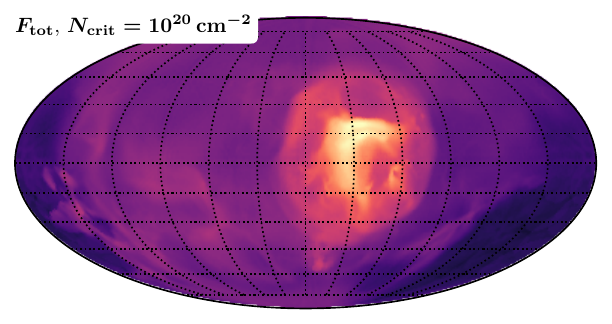}
    \includegraphics[width=0.32\textwidth]{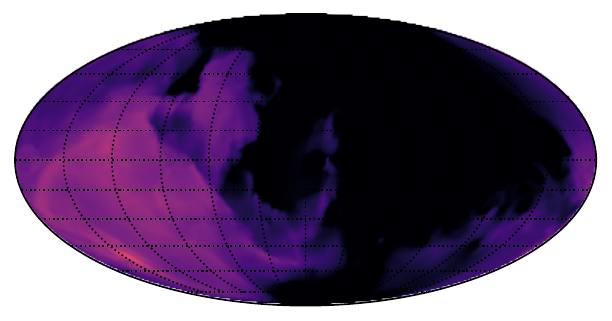}
    \includegraphics[width=0.32\textwidth]{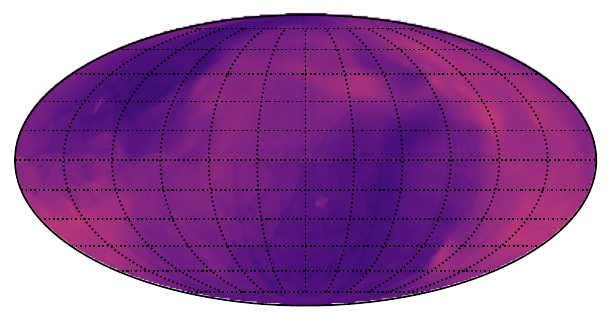}\\
    \includegraphics[width=0.32\textwidth]{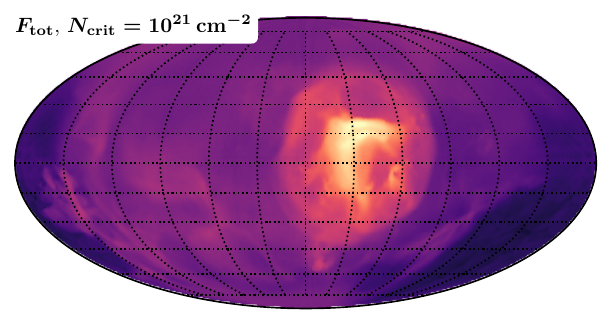}
    \includegraphics[width=0.32\textwidth]{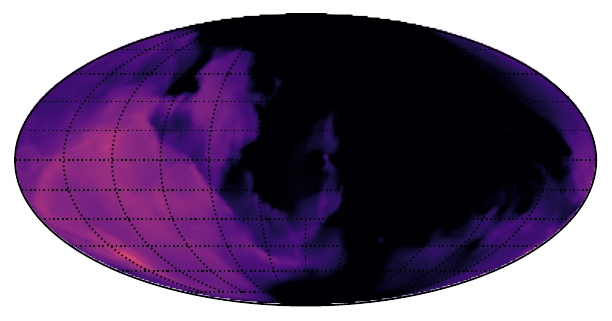}
    \includegraphics[width=0.32\textwidth]{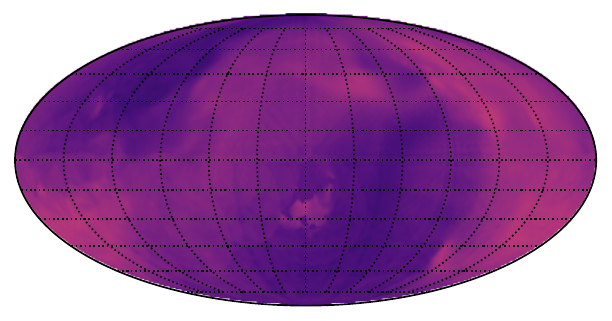}\\
    \includegraphics[width=0.9\textwidth]{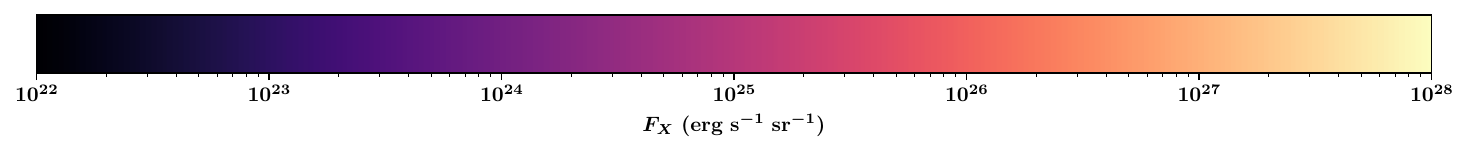}\\
    \includegraphics[width=0.32\textwidth]{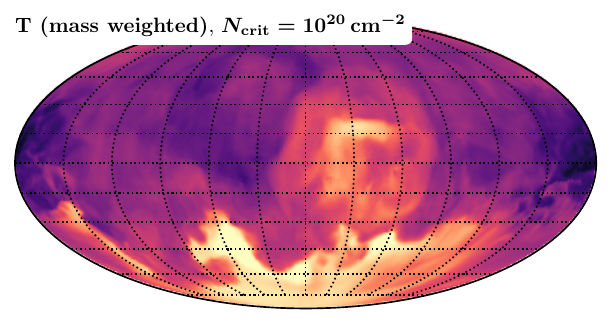}
    \includegraphics[width=0.32\textwidth]{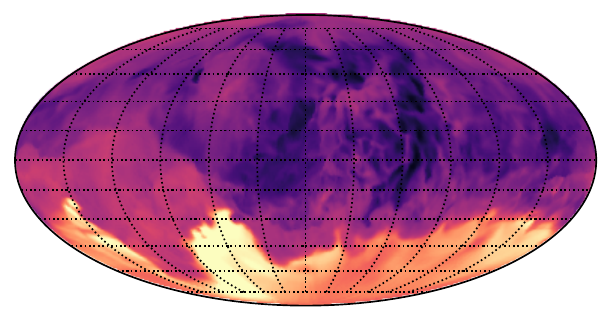}
    \includegraphics[width=0.32\textwidth]{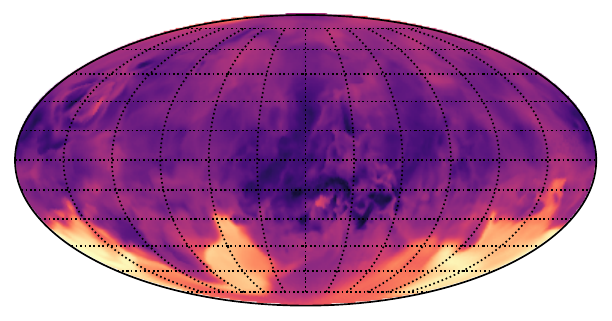}\\
    \includegraphics[width=0.32\textwidth]{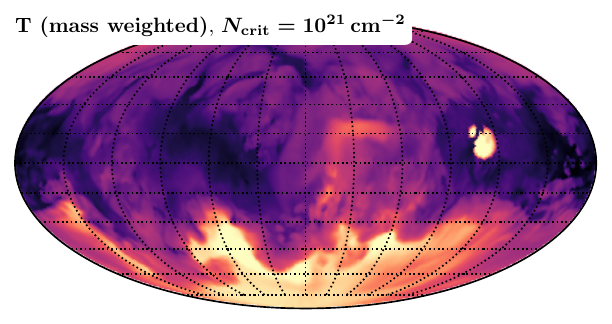}
    \includegraphics[width=0.32\textwidth]{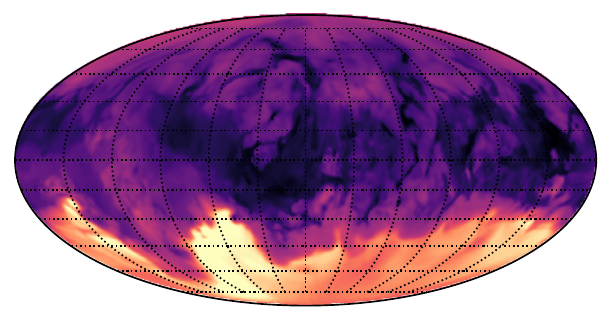}
    \includegraphics[width=0.32\textwidth]{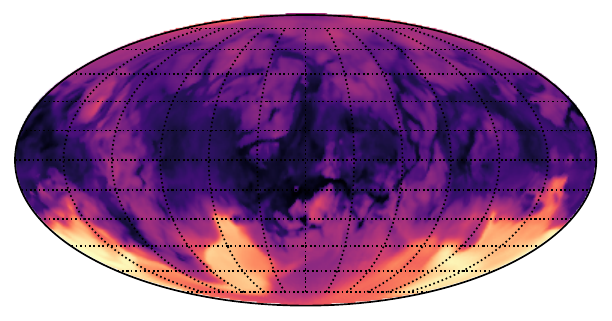}\\
    \includegraphics[width=0.32\textwidth]{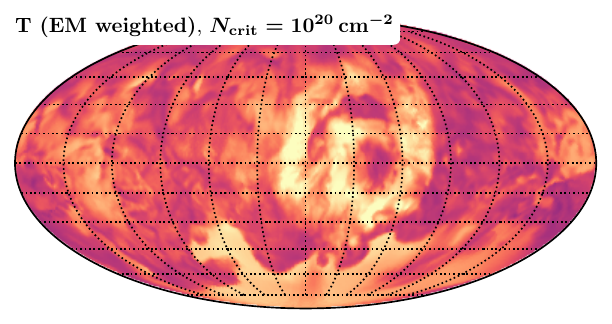}
    \includegraphics[width=0.32\textwidth]{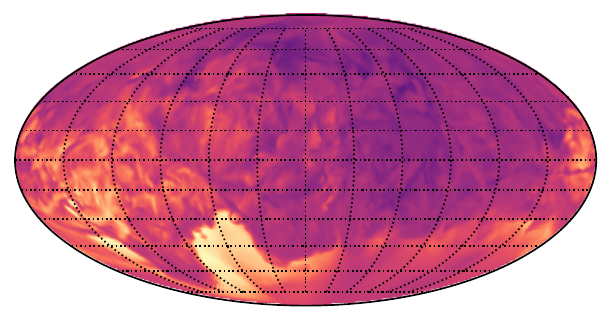}
    \includegraphics[width=0.32\textwidth]{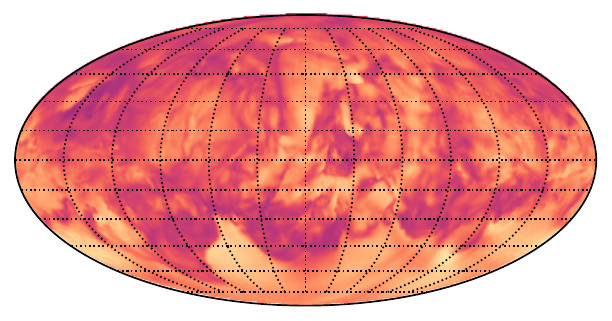}
    \includegraphics[width=0.32\textwidth]{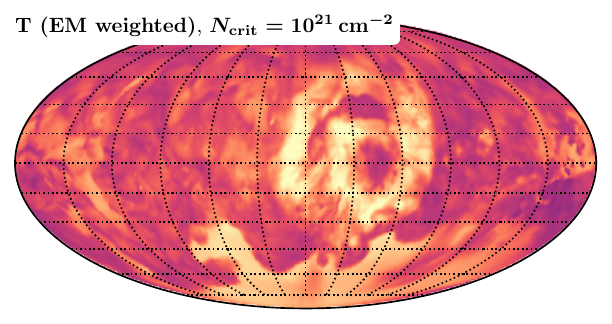}
    \includegraphics[width=0.32\textwidth]{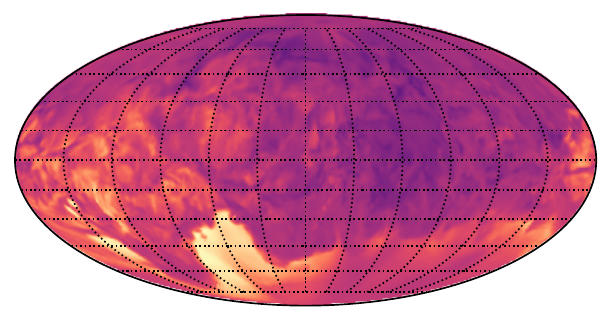}
    \includegraphics[width=0.32\textwidth]{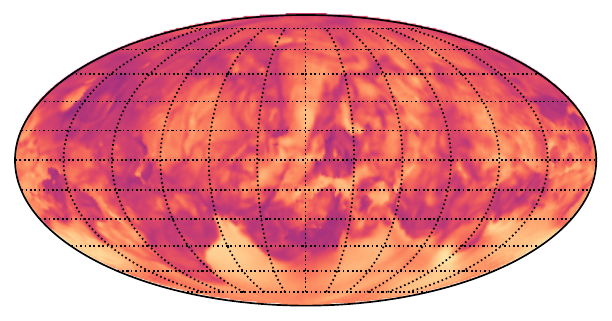}
    \includegraphics[width=0.9\textwidth]{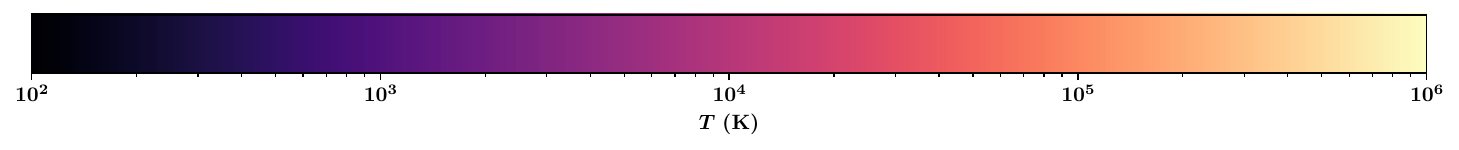}
    \caption{Comparison of all-sky maps for different critical column densities. From left to right we show again the active and two quiescent states. From top to bottom we compare the total flux (first two rows), the mass weighted (third and fourth row) and the EM weighted temperature (two two rows) for $N_\mathrm{crit}=10^{20}\,\mathrm{cm}^{-2}$ and $N_\mathrm{crit}=10^{21}\,\mathrm{cm}^{-2}$ (first vs. second row in each group). For the total flux and the EM weighed temperature, the differences are small and only visible for small patches of the sky. Only the mass weighted temperature shows a systematic difference in the region around the equator, which reflects the deeper integration into the dense and cold shell.}
    \label{fig:Ncrit-maps}
\end{figure*}

We highlight that the critical column density only has minor effects on most of the all-sky maps we investigate. This is illustrated in Fig.~\ref{fig:Ncrit-maps}. From left to right we again show the active and two quiescent states. From top to bottom we depict the total X-ray flux (top two rows), the mass weighted temperature (row three and four) as well as the EM weighted $T$ (two bottom rows). By far the strongest differences are seen in the temperature maps, in particular in the mass weighted one. This is expected since a larger integration depth into the bubble shell probes denser and thus colder gas. For all other quantities the differences are located in small patches of the sky. None of the large-scale features nor the overall scaling of the values is affected.

\section{Examination of Closest Cells}
\label{app:closest-cell}

\begin{figure}
\includegraphics[width=\linewidth]{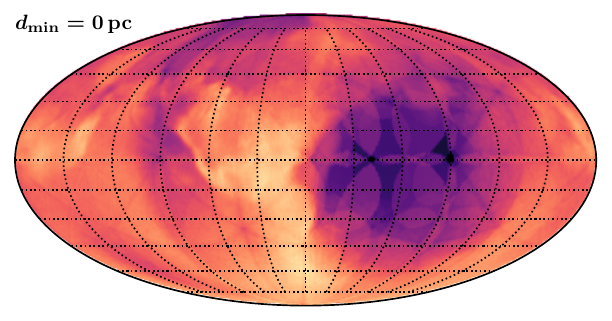}\\
\includegraphics[width=\linewidth]{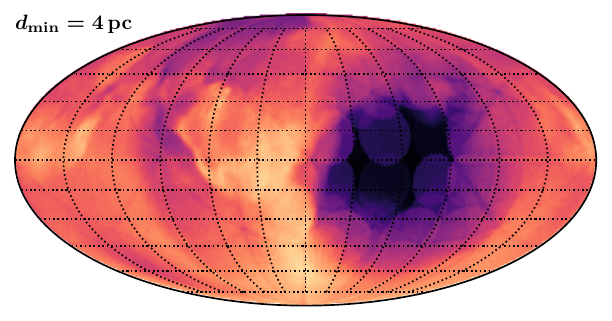}
\includegraphics[width=\linewidth]{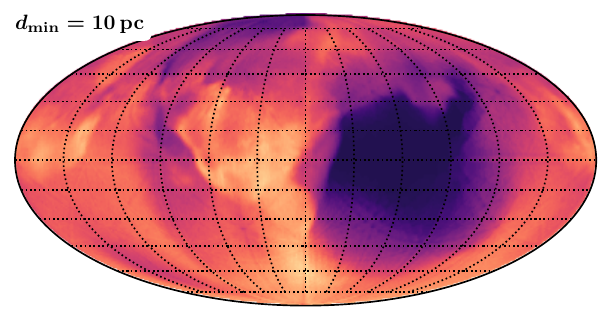}\\
\includegraphics[width=\linewidth]{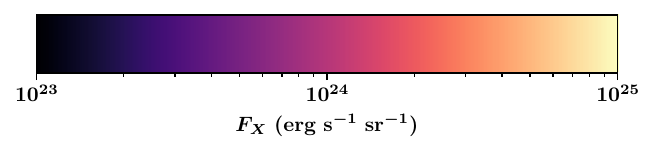}
\caption{Impact of the closest cells on the all-sky flux maps. Shown are from top to bottom flux maps with $d_\mathrm{min}=0, 4$, and $10\,\mathrm{pc}$. We note that without limiting the distance, close cells lead to grid features in the projection.}
\label{fig:closest-cells}
\end{figure}

While making Mollweide projections of X-ray luminosity for each epoch, we notice that in some cases the signal is dominated by cells that lie very close to the observer and therefore cover a disproportionately large field of view. To address this issue, a minimum distance parameter was introduced to reduce the impact of nearby cells. This parameter, $d_\mathrm{min}$, was carefully set to 10~pc after extensive testing to spread the radiation from nearby cells over a broader area. We illustrate the effect of very close cells in Fig.~\ref{fig:closest-cells} for the first analysis snapshot at $8\,\mathrm{Myr}$ for 0, 4 and 10~pc minimum radius.

\end{appendix}

\end{document}